\documentclass[12pt]{article}
\usepackage{amsmath}
\usepackage{amssymb}
\usepackage{slashed}
\usepackage{graphicx}
\usepackage{subfigure}
\usepackage{cite}
\usepackage{url}
\usepackage[small]{caption}
\begin{document}
\baselineskip 0.6cm

\def\bra#1{\langle #1 |}
\def\ket#1{| #1 \rangle}
\def\inner#1#2{\langle #1 | #2 \rangle}

\begin{titlepage}

\begin{flushright}
UCB-PTH-15/08\\
\end{flushright}

\vskip 1.8cm

\begin{center}
{\Large \bf Flat-space Quantum Gravity in AdS/CFT}

\vskip 0.7cm

{\large Yasunori Nomura, Fabio Sanches, and Sean J. Weinberg}

\vskip 0.4cm

{\it Berkeley Center for Theoretical Physics, Department of Physics,\\
 University of California, Berkeley, CA 94720, USA}

\vskip 0.1cm

{\it Theoretical Physics Group, Lawrence Berkeley National Laboratory,
 CA 94720, USA}

\vskip 0.8cm

\abstract{Motivated by the task of understanding microscopic dynamics 
of an evolving black hole, we present a scheme describing gauge-fixed 
continuous time evolution of quantum gravitational processes in 
asymptotically flat spacetime using the algebra of CFT operators. 
This allows us to study the microscopic dynamics of the Hawking emission 
process, although obtaining a full $S$-matrix may require a modification 
of the minimal scheme.  The role of the operator product expansion 
is to physically interpret the resulting time evolution by decomposing 
the Hilbert space of the states for the entire system into those 
for smaller subsystems.  We translate the picture of an evaporating 
black hole previously proposed by the authors into predictions for 
nonperturbative properties of the CFTs that have weakly coupled dual 
gravitational descriptions.  We also discuss a possible relationship 
between the present scheme and a reference frame change in the bulk.}

\end{center}
\end{titlepage}

\section{Introduction}
\label{sec:intro}

Quantum gravity has been elusive so far.  Our lack of understanding 
manifests itself especially prominently in the study of black hole physics, 
where basic questions such as unitarity of the evolution and the smoothness 
of horizons are still under debate~\cite{Hawking:1976ra,'tHooft:1990fr,%
Almheiri:2012rt}.  Ever since the discovery of the thermodynamic behavior 
of black holes~\cite{Bekenstein:1973ur,Bardeen:1973gs,Hawking:1974rv}, 
we have been searching for the deeper structure of spacetime and gravity 
beyond that described by general relativity.  In fact, we may need to 
revise the concept of spacetime itself, as suggested by the holographic 
principle~\cite{'tHooft:1993gx,Susskind:1994vu} and complementarity 
hypothesis~\cite{Susskind:1993if,Susskind:2005js}.  Furthermore, it 
seems that a perturbative approach to gravity is incapable of revealing 
the real nature of spacetime.

AdS/CFT duality~\cite{Maldacena:1997re,Gubser:1998bc} provides a possible 
approach to quantum gravity at the nonperturbative level, albeit in 
spacetimes that are asymptotically AdS.  Motivated by the task of 
understanding microscopic dynamics of an evolving black hole, the 
first half of this paper presents a scheme which can describe quantum 
gravitational processes in asymptotically flat spacetime using CFTs. 
A key point is that this does not require a true (holographic) theory 
of quantum gravity in asymptotically flat spacetime, i.e.\ one accommodating 
the full Bondi-Metzner-Sachs symmetry at null infinity~\cite{Bondi:1962px} 
and giving a fully unitary $S$-matrix in the asymptotically flat spacetime. 
Instead, we can describe flat-space quantum gravitational processes 
by focusing on time scales sufficiently shorter than the AdS time 
scale and in a sufficiently central region of the global AdS space 
in a single AdS volume at the center.  This description can be made 
extremely (and perhaps infinitely) accurate by making the AdS length 
scale large compared to the scale of interest.

The relevant Heisenberg-picture states in the gravitational bulk 
are represented by CFT operators at the point corresponding to the 
infinite past, which exists on the flat Euclidean space obtained by 
conformally compactifying the boundary spacetime in which the CFT 
originally lived.  We need not use the concept of CFT fields:\ knowing 
the spectrum and algebra of these operators is enough to understand the 
dynamics of our interest.  In particular, by identifying the dilatation 
$D$ of the conformal symmetry with time translation in the bulk, we 
can describe continuous time evolution (not just an $S$-matrix type 
quantity) in the gravitational bulk.  Since the CFT description 
eliminates all the gauge redundancies in the gravitational theory 
(including those associated with the holographic reduction of degrees 
of freedom), this provides a fully gauge-fixed description of physical 
observables in quantum gravity.  Our particular choice of identifying 
$D$ as time translation corresponds to taking the reference clock 
to be in the asymptotic region~\cite{DeWitt:1967yk}.

Despite its conceptual simplicity, current theoretical technology does 
not allow us to compute the physics of black holes explicitly by following 
the above program.  In the second half of this paper, we therefore 
adopt a different strategy and use information from the gravitational 
description to study what properties the dual CFTs must possess.  In 
particular, we describe how the picture of an evaporating black hole 
in Refs.~\cite{Nomura:2014woa,Nomura:2014voa}, proposed to solve the 
black hole information problem~\cite{Hawking:1976ra,Almheiri:2012rt}, 
is translated into the CFT language given here.  This has at least 
two virtues.  First, since the physics of black holes is expected to 
be universal, the structures we identify can be viewed as predictions 
for nonperturbative properties of the CFTs that have weakly coupled 
gravitational descriptions.  In principle, this allows us to test 
(aspects of) the picture of Refs.~\cite{Nomura:2014woa,Nomura:2014voa}, 
perhaps with some future theoretical developments.  Second, the translated 
CFT description clarifies the concept of spacetime-matter duality 
introduced in Refs.~\cite{Nomura:2014woa,Nomura:2014voa}:\ the black 
hole microstates play roles of both spacetime and matter, but in fact 
are neither.  In the CFT language, this can be stated as properties 
exhibited by the operators corresponding to black hole microstates.

The emergence of spacetime and gravity in AdS/CFT is an important 
subject, and it has been studied by many authors from various different 
angles, e.g., in Refs.~\cite{Balasubramanian:1998sn,Polchinski:1999ry,%
Hamilton:2006az,Gary:2009ae,Heemskerk:2009pn,Fitzpatrick:2010zm,%
Penedones:2010ue,ElShowk:2011ag,Kabat:2011rz,Heemskerk:2012mn,%
Fitzpatrick:2012cg,Kabat:2013wga,Fitzpatrick:2014vua,Almheiri:2014lwa,%
Bao:2015bfa,Verlinde:2015qfa}.  Our analysis builds on many of these 
works, which we refer to more explicitly as we go along.  It asserts 
that when a relevant CFT possesses a finite central charge, which is 
necessary to describe nonperturbative gravitational processes in the 
bulk, the sector allowing for a semiclassical particle interpretation 
comprises only a tiny subset of the whole degrees of freedom representing 
the single AdS volume.  This elucidates why, in contrast with what 
is postulated in Refs.~\cite{Almheiri:2012rt,Almheiri:2013hfa,%
Marolf:2013dba,Polchinski:2015cea}, the microscopic information about 
a black hole cannot be viewed as propagating in semiclassical spacetime 
in the near black hole region.

The organization of this paper is as follows.  In Section~\ref{sec:flat-AdS}, 
we present the scheme which describes flat-space quantum gravitational 
processes using CFTs.  We discuss under what conditions and to what 
extent the CFTs may provide such descriptions and how the continuous 
time evolution picture in the gravitational bulk arises from the algebra 
of CFT operators defined at a point in Euclidean spacetime.  We then 
discuss in Section~\ref{sec:BH-AdS} how these quantum gravitational 
processes may be physically interpreted (purely) in CFTs.  This requires 
us to decompose operators into smaller elements, which corresponds to 
decomposing the Hilbert space of the states for the ``entire universe'' 
into those for smaller subsystems.  We discuss how this can be done using 
the operator product expansion (OPE), defined as the action of operators 
on general states.  We finally apply the scheme to black hole states 
to see how the picture of Refs.~\cite{Nomura:2014woa,Nomura:2014voa} 
may be realized in CFTs.  Section~\ref{sec:discuss} is devoted to 
the summary and discussion.

\section{Embedding Asymptotically Flat Spacetime in AdS/CFT}
\label{sec:flat-AdS}

One goal of this paper is to use AdS/CFT duality to study how the 
structure of (a class of) CFTs encodes the physics of an evaporating, 
dynamically formed black hole.  As a first step, here we ask how 
the physics of flat-space quantum gravity may be encoded in CFTs.

Throughout the paper, we focus on the class of CFTs whose dual descriptions 
possess energy intervals in which physics is well described by weakly 
coupled effective field theories with Einstein gravity in spacetimes 
one dimension higher than those of the corresponding CFTs.  Specifically, 
for a $d$-dimensional CFT, we require that the dual theory has
\begin{equation}
  l_{\rm P} < l_{\rm s} < R_{\rm c} \ll R,
\label{eq:hierarchy}
\end{equation}
where $l_{\rm P}$, $l_{\rm s}$, and $R$ are the $(d+1)$-dimensional 
Planck length, string length, and AdS curvature length, respectively, 
and $R_{\rm c}$ collectively represents characteristic length scales 
associated with compact extra dimensions beyond $d+1$ spacetime dimensions. 
The precise and general conditions for a CFT to give such descriptions 
are not yet understood, although some necessary and/or sufficient conditions 
to have weakly coupled dual gravitational descriptions have been discussed 
in various contexts~\cite{Heemskerk:2009pn,ElShowk:2011ag,Fitzpatrick:2012cg,%
Bao:2015bfa}.  In this paper, we assume the existence of the relevant CFTs.

Since the flat-space limit corresponds to length scales smaller than 
the AdS radius $R$, we are interested only in a single AdS volume.  We 
choose to work in global AdS spacetime, focusing on a single AdS volume 
at the center.  This has the virtue that we need not be concerned with 
the Poincar\'{e} horizon in dual AdS$_{d+1}$ descriptions.  In particular, 
we take our CFTs to live in ${\bf S}^{d-1} \times {\bf R}$, where 
${\bf S}^{d-1}$ represents the $(d-1)$-dimensional sphere with 
radius $R$.

\subsection{AdS radius as an IR and UV cutoff}
\label{subsec:R_cutoff}

AdS/CFT duality is believed to hold between a $d$-dimensional CFT 
and string theory on a spacetime whose asymptotic region contains an 
AdS$_{d+1}$ factor.  Suppose the spacetime geometry in the string theory 
side is asymptotically AdS$_{d+1} \times X$, where $X$ is a compact 
space.  It seems possible to make $X$ as small as of order the AdS 
radius $R$.  This was indeed the case in originally discussed examples 
of AdS/CFT~\cite{Maldacena:1997re}, and we do not anticipate any 
obstacles keeping $X$ this size (or smaller) in more elaborate 
setups, for example in nonsupersymmetric cases.  This allows us to 
dimensionally reduce on $X$, yielding theory in AdS$_{d+1}$ that 
contains Kaluza-Klein towers associated with $X$.  In this subsection 
we work in this framework.

\begin{figure}[t]
\begin{center}
  \includegraphics[height=7cm]{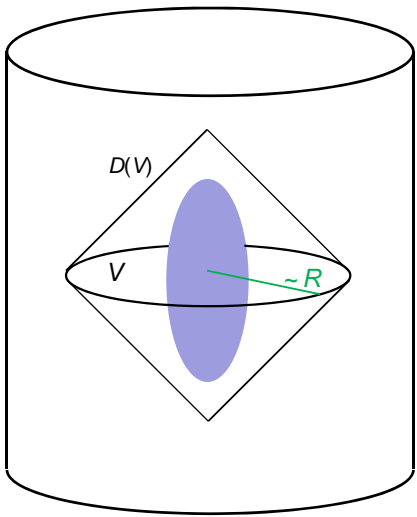}
\end{center}
\caption{A single AdS volume $V$ is selected at the center of the global 
 AdS space at some reference time $t_0$.  We are concerned with processes 
 occurring in a sufficiently inner region of the domain of dependence 
 of $V$, $D(V)$.}
\label{fig:center}
\end{figure}

Let us consider a single AdS volume $V$ at the center of the bulk at 
some time $\tau = t_0$:
\begin{equation}
  0 \leq r < R,
\label{eq:single_V}
\end{equation}
where $\tau$ and $r$ are the AdS time and radial coordinates defined 
by the metric
\begin{equation}
  ds^2 = -\biggl( 1+\frac{r^2}{R^2} \biggr)\, d\tau^2 
    + \frac{dr^2}{1+\frac{r^2}{R^2}} + r^2 d\Omega^2.
\label{eq:def-r}
\end{equation}
This region serves as our proxy of (an equal-time hypersurface of) flat 
space, in which $R$ plays the role of the IR cutoff.  When we refer 
to flat-space physics, we mean physical processes occurring in (a 
sufficiently interior part of) this region; see Fig.~\ref{fig:center}. 
In general, excitations involved in these processes modify the metric 
in Eq.~(\ref{eq:def-r}), in which case the choice of time slice, 
$\tau = t_0$, refers to that at the boundary of the region, $\partial V$.%
\footnote{This statement becomes fully unambiguous only in the true 
 flat-space limit, in which we focus on processes occurring at the center, 
 $0 \leq r < \epsilon R$ with $\epsilon \rightarrow 0$.  Naively, this 
 limit may be taken by keeping the number of possible states, $N_\epsilon$, 
 realized in this region large and finite.  Here, $\ln N_\epsilon \sim 
 \epsilon^{d-1} c$ with $c$ being the central charge of the dual CFT, 
 so that $c \rightarrow \infty$; see Eq.~(\ref{eq:holo-CFT}).  (This 
 corresponds to taking the $\epsilon \rightarrow 0$ limit keeping 
 $\epsilon R/l_{\rm P}$ large and finite.)  To discuss physics of black 
 holes, however, we need to analyze CFT operators with dimensions scaling 
 as positive powers in $c$, which prevents us from taking $c$ to be 
 literally infinite (see Section~\ref{subsec:dual}).  We thus take 
 $\epsilon \sim O(1)$ ($< 1$) in this paper.  Any ambiguities arising 
 from the prescription of dealing with $\partial V$ only affect physics 
 at the IR cutoff scale (which may be made arbitrarily small by making 
 $\epsilon$ smaller). \label{ft:IR}}

What are the CFT operators representing physical configurations in 
the volume $V$ (or more precisely, in the region inside $\partial V$)? 
Imagine that the CFT is defined on a flat $d$-dimensional Euclidean 
spacetime obtained by performing a Weyl transformation on (Euclideanized) 
${\bf S}^{d-1} \times {\bf R}$ and adding a point, $x=x_{-\infty}$, 
corresponding to $\tau = -\infty$.  Here, $x$ represents the coordinates 
of the $d$-dimensional space in which the CFT lives.  We may then define 
the set of (gauge-invariant) operators acting at this point, $\Psi$'s, 
which corresponds to all the elements of the Hilbert space of this CFT 
in radial quantization.  The CFT state created by any of these operators, 
$\Psi \ket{0}_{\rm CFT}$, then corresponds to a Heisenberg state in 
the gravitational theory, representing a full spacetime history in the 
bulk.  We can then ask what subset of these CFT operators provides a 
complete basis $\{ \Psi_{A=1,2,\cdots} \}$ for the bulk states that 
can be interpreted as having excitations only in the region inside 
$\partial V$ at time $t_0$.

Let us recall that AdS/CFT relates the central charge $c$ of the CFT 
with the AdS radius and the $(d+1)$-dimensional Planck length as
\begin{equation}
  c \sim \left( \frac{R}{l_{\rm P}} \right)^{d-1}.
\label{eq:c-dual}
\end{equation}
Since we are interested in physics in (large) flat spacetime, we take 
$R \gg l_{\rm P}$.  Namely, we are considering CFTs with
\begin{equation}
  c \gg 1.
\label{eq:large-c}
\end{equation}
In general, a CFT operator $\Psi$ is given by a superposition of primary 
and descendant operators ${\cal O}_I$:%
\footnote{Note that the index $I$ in general involves spacetime indices 
 in the (Euclidean) $d$-dimensional space.}
\begin{equation}
  \Psi = \sum_I \alpha_I\, {\cal O}_I.
\label{eq:Psi_x}
\end{equation}
Here, ${\cal O}_I$'s are defined by the dilatation $D$ and special conformal 
transformations $K^\mu$ whose center is at $x=x_{-\infty}$, and we take 
their normalization such that the states created by them are appropriately 
normalized in Lorentzian ${\bf S}^{d-1} \times {\bf R}$:
\begin{equation}
  \bra{0} {\cal O}_I^\dagger {\cal O}_J \ket{0} = \delta_{IJ},
\label{eq:Oi-norm}
\end{equation}
where $\ket{0}$ is the CFT vacuum state.  (This requires unconventional 
relative normalizations between primary and descendant operators within 
a single conformal multiplet.%
\footnote{For example, if a scalar primary operator ${\cal O}^{(0)}$ of 
 dimension $\Delta$ is normalized such that $\bra{0} {\cal O}^{(0)\dagger} 
 {\cal O}^{(0)} \ket{0} = 1$, the conventionally defined descendant 
 operators ${\cal O}^{(1)}_\mu = [ P_\mu, {\cal O}^{(0)}]$ and 
 ${\cal O}^{(n)}_{\mu_1 \cdots \mu_n} = [ P_{\mu_1}, {\cal O}^{(n-1)}_{\mu_2 
 \cdots \mu_n}]$ ($n = 2,3,\cdots$) give $\bra{0} {\cal O}^{(1)\dagger}_\mu 
 {\cal O}^{(1)}_\rho \ket{0} = 2\Delta \eta_{\mu\rho}$, $\bra{0} 
 {\cal O}^{(2)\dagger}_{\mu\nu} {\cal O}^{(2)}_{\rho\sigma} \ket{0} 
 = 4\Delta(\Delta+1) (\eta_{\mu\rho} \eta_{\nu\sigma} + \eta_{\mu\sigma} 
 \eta_{\nu\rho}) - 4\Delta \eta_{\mu\nu} \eta_{\rho\sigma}$, and so on.}%
)  According to AdS/CFT, the dimension $\Delta$ of a CFT (primary or 
descendant) operator ${\cal O}$ is related with the bulk quantities as
\begin{equation}
  \Delta \sim ER,
\label{eq:Delta-dual}
\end{equation}
where $E$ is the energy of the bulk state corresponding to ${\cal O}$ 
as measured with respect to $\tau$.

The relations in Eqs.~(\ref{eq:c-dual},~\ref{eq:Delta-dual}) imply 
that a CFT operator $\Psi$ containing a primary or descendant operator 
with $\Delta > c$ corresponds to a bulk state involving a component with 
$E > R^{d-2}/l_{\rm P}^{d-1}$.  Such an operator cannot be an element 
of the basis $\{ \Psi_A \}$ we look for, since energies this large 
cannot fit into $V$---they would lead to large black holes, whose 
Schwarzschild radii are larger than $R$.  A basis element of 
$\{ \Psi_A \}$ thus can be written as
\begin{equation}
  \Psi_A = \sum_{I \in \{ I| \Delta_I < c \} }\!\! 
    \alpha_{A,I}\, {\cal O}_I,
\label{eq:Psi_A}
\end{equation}
where
\begin{equation}
  \sum_I \alpha_{A,I}^* \alpha_{B,I} = \delta_{AB}.
\label{eq:alpha-norm}
\end{equation}
Note that the coefficients $\alpha_{A,I}$ depend on $t_0$, and 
$\alpha_{A,I}^*$ refers to the complex conjugate of $\alpha_{A,I}$ 
in Lorentzian ${\bf S}^{d-1} \times {\bf R}$.

The set of CFT operators $\Psi_A$ is smaller than the set spanned by 
all possible independent linear combinations of ${\cal O}_I$'s with 
$\Delta_I < c$.  This is because some of these linear combinations 
correspond to bulk states in which not all the excitations are confined 
within $\partial V$ at time $t_0$.  While an explicit expression for the 
sets of $\alpha_{A,I}$'s giving correct basis operators, $\Psi_A$, is 
not known, they must be uniquely determined (up to basis changes and IR 
ambiguities discussed in footnote~\ref{ft:IR}) once the theory is fixed 
and time $t_0$ is chosen, since the CFT construction here selects time 
slicing in the bulk.  Note that the choice of $t_0$ is simply a convention 
for how we embed our ``flat spacetime'' into the AdS spacetime.  We refer 
to the set of basis operators determined in this way as ${\cal B}$:
\begin{equation}
  {\cal B} = \{ \Psi_A \}.
\label{eq:set-B}
\end{equation}
This set contains complete information about physics occurring in our 
``flat spacetime,'' i.e.\ (a sufficiently small part of) the domain of 
dependence of $V$, $D(V)$ (see Fig.~\ref{fig:center}).

We stress that the dimensions of operators comprising $\Psi_A$'s are 
bounded both from below (by the unitarity bound) and above (by $c$); 
namely, the AdS radius provides a cutoff both in the IR and UV.  In 
fact, for finite $R$ (in units of $l_{\rm P}$), the number of elements 
of ${\cal B}$, $N_{\cal B}$, is finite---the holographic principle 
states that~\cite{'tHooft:1993gx,Susskind:1994vu,Bousso:1999xy} 
\begin{equation}
  \ln N_{\cal B} 
  = \frac{\Omega_{d-1}}{4} \left( \frac{R}{l_{\rm P}} \right)^{d-1} 
  \sim c,
\label{eq:holo-CFT}
\end{equation}
where $\Omega_{d-1} = 2\pi^{d/2}/\Gamma(d/2)$ is the volume of the unit 
$(d-1)$-sphere, and  we have used Eq.~(\ref{eq:c-dual}).%
\footnote{When we make $\epsilon$ discussed in footnote~\ref{ft:IR} 
 smaller, we are focusing on a smaller subset of the basis operators 
 ${\cal B}_\epsilon \subset {\cal B}$.  The number of elements in 
 this subset, $N_\epsilon$, is given by $\ln N_\epsilon \sim 
 \epsilon^{d-1} c$.}
This is the CFT statement for holography in flat spacetime, in which (unlike 
a large AdS region) a volume does not scale as the area surrounding it.

\subsection{Flat-space quantum gravity in {\boldmath $d+1$} dimensions}
\label{subsec:dual}

We now discuss in more detail how we may extract physics of flat-space 
quantum gravity.  We begin by considering the (potentially hypothetical) 
situation in which the compact space $X$ can be completely ignored. 
Specifically, we consider the case in which the gravitational theory 
contains only two length scales
\begin{equation}
  \tilde{l}_{\rm P} \sim l_{\rm s} \sim R_{\rm c} \ll R.
\label{eq:2-scales}
\end{equation}
Here, $\tilde{l}_{\rm P}$ is the Planck length in AdS$_{d+1} \times X$, 
which is related with the $(d+1)$-dimensional Planck length, $l_{\rm P}$, 
by
\begin{equation}
  l_{\rm P}^{d-1} = \frac{\tilde{l}_{\rm P}^{d+n-1}}{V_X},
\label{eq:dim-red}
\end{equation}
where $n$ and $V_X$ are the dimension and volume of $X$, respectively. 
(Here we take $V_X = R_{\rm c}^n \sim l_{\rm s}^n$, so that 
$\tilde{l}_{\rm P} \sim l_{\rm P}$.)  As we will discuss below, 
this setup might not be realized in a consistent theory of quantum 
gravity (and so as a dual description of a CFT).  However, it provides 
a useful starting point for our discussion.

Suppose we take the limit in which the cutoff for flat-space 
physics is removed, $R/l_{\rm P} \rightarrow \infty$, which 
corresponds to $c \rightarrow \infty$.  The relations in 
Eqs.~(\ref{eq:c-dual},~\ref{eq:Delta-dual}) then imply that 
trans-Planckian physics in the bulk is encoded in the structure 
of operators with extremely large (infinite) dimensions
\begin{equation}
  \Delta \gtrsim c^{\frac{1}{d-1}}.
\label{eq:Delta-1}
\end{equation}
As we have seen before, operators with $\Delta \gtrsim c$ do not correspond 
to physics in flat spacetime---they represent intrinsically AdS physics. 
This implies that physics of flat-space black holes is encoded in 
operators with%
\footnote{If we make $\epsilon$ smaller (see footnote~\ref{ft:IR}), 
 the upper bound on the operator dimensions becomes $\epsilon^{d-2} c$, 
 instead of $c$, since we are not interested in energies leading to 
 black holes whose Schwarzschild radii are larger than $\epsilon R$. 
 The same also applies to Eqs.~(\ref{eq:c-range},~\ref{eq:c-range_ex}), 
 implying that $\epsilon$ cannot be made of order $(g_{\rm s}^{d-n+1} 
 / c^{n-1})^{1/dn}$ or smaller if the size of $X$ indeed takes the 
 values discussed there.}
\begin{equation}
  c^{\frac{1}{d-1}} \lesssim \Delta \lesssim c.
\label{eq:BH-naive}
\end{equation}
In the rest of this paper, we assume $d > 2$.

A main point of the above discussion is that to study physics of 
flat-space quantum gravity, in particular that of black holes, we 
need to analyze the structure of operators that have very large dimensions, 
scaling as positive powers in $c$.  The detailed range of operator 
dimensions on which we need to focus, however, depends on the size 
(and shape) of the extra dimensions as well as the physics we are 
interested in.  Below, we discuss the effect of extra dimensions on 
this issue.  In particular, we discuss what strategy one may adopt 
in studying flat-space black holes using AdS/CFT.

Let us ask how small a compact space $X$ can actually be.  From the 
viewpoint of effective AdS$_{d+1}$ field theory, there is no restriction 
on the size of $X$.  In particular, there is nothing wrong in taking 
$V_X \sim l_{\rm s}^n$ so that Eq.~(\ref{eq:2-scales}) is satisfied. 
However, nonperturbative effects of quantum gravity may impose constraints 
on the size and shape of $X$.  For example, Ref.~\cite{ArkaniHamed:2006dz} 
argues that for $d=4$ and $n = 5$ the volume of $X$ must satisfy 
$(V_X/l_{\rm s}^5) \gtrsim g_{\rm s} (R/l_{\rm s})$, where $g_{\rm s}$ 
is the string coupling.  A naive extension of this to arbitrary $d$ 
and $n$ gives%
\footnote{The validity of this extension beyond $d=4$ is not clear, 
 since we have not taken into account nontrivial profiles of the 
 dilaton field induced by relevant branes.  Here we present it simply 
 to illustrate the basic issue.}
\begin{equation}
  \frac{V_X}{l_{\rm s}^n} \gtrsim g_{\rm s} \frac{R}{l_{\rm s}}.
\label{eq:WGC}
\end{equation}
If bounds like this indeed exist, we will not be able to have a consistent 
setup in which the effect of extra dimensions can be completely neglected.

\begin{figure}[t]
\begin{center}
  \includegraphics[height=8.5cm]{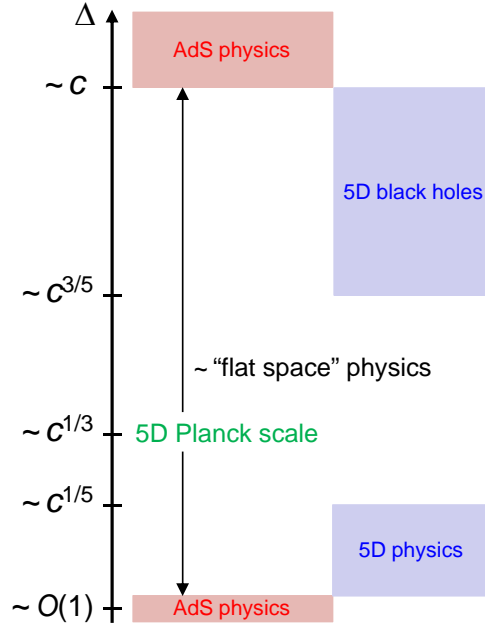}
\end{center}
\caption{For AdS$_5 \times X_5$, the 5-dimensional Planck scale in the bulk 
 corresponds to the scaling dimension of order $c^{1/3}$ in the 4-dimensional 
 dual CFT.  If there is a lower bound on the size of $X_5$ as suggested 
 by the argument in Ref.~\cite{ArkaniHamed:2006dz}, then (assuming $X_5$ 
 has only a single scale) the physics of 5-dimensional black holes can be 
 explored by considering operators with dimensions between $\sim c^{3/5}$ 
 and $\sim c$.  (To study 5-dimensional perturbative processes, we may 
 focus on operators with dimensions $\lesssim c^{1/5}$.)}
\label{fig:Delta}
\end{figure}

How do we study flat-space black holes in such cases?  One way would 
be to consider a setup in which the extra dimensions are large, 
e.g.\ of order the AdS radius $R$, and investigate $(d+n+1)$-dimensional 
physics at length scales smaller than $R$.  This, however, might allow 
us to study only special classes of black holes, e.g.\ those in 10- 
or 11-dimensional spacetimes with a large number of supersymmetries 
(see, e.g., Ref.~\cite{Asplund:2008xd}).  Another possibility is to 
make $X$ as small as possible and study black holes larger than $X$. 
Imagine that $X$ can indeed saturate the bound in Eq.~(\ref{eq:WGC}) 
with all the length scales being roughly comparable.  The compactification 
length $R_{\rm c} \sim V_X^{1/n}$ is then given by
\begin{equation}
  R_{\rm c} \sim \left( \frac{g_{\rm s}^{d-n+1}}{c^{n-1}} 
    \right)^{\frac{1}{dn}} R.
\label{eq:R_c}
\end{equation}
The Schwarzschild radius of a black hole of mass $M$ is larger than 
$R_{\rm c}$ if
\begin{equation}
  M \gtrsim M_{\rm min} \sim \left( g_{\rm s}^{(d-2)(d-n+1)} 
    c^{d+2n-2} \right)^{\frac{1}{dn}} \frac{1}{R}.
\label{eq:M}
\end{equation}
This implies that we can study physics of flat-space black holes in 
$(d+1)$-dimensional spacetime by analyzing CFT operators with the 
dimensions $\Delta$ satisfying
\begin{equation}
  M_{\rm min} R \lesssim \Delta \lesssim c.
\label{eq:c-range}
\end{equation}
For $d=4$ and $n=5$, for example, we may investigate CFT operators with
\begin{equation}
  c^{\frac{3}{5}} \lesssim \Delta \lesssim c,
\label{eq:c-range_ex}
\end{equation}
to study 5-dimensional flat-space black holes; see Fig.~\ref{fig:Delta}. 

Bounds on $R_{\rm c}$ as the ones described above, if they indeed exist, 
provide a restriction on how much our minimal formalism in the previous 
subsection can capture flat-space physics.  Consider a black hole of mass 
$M$ in AdS$_{d+1}$, whose Schwarzschild radius is smaller than $R$ (so that 
it behaves as a flat-space black hole).  It has the Hawking temperature
\begin{equation}
  T_{\rm H}(M) = \frac{d-2}{4\pi} \biggl[ 
    \frac{(d-1)\Omega_{d-1}}{16\pi} \biggr]^{\frac{1}{d-2}} 
    \frac{1}{(M l_{\rm P}^{d-1})^{\frac{1}{d-2}}},
\label{eq:T_H}
\end{equation}
so that a black hole with an initial mass $M_0$ has the lifetime 
of order
\begin{equation}
  \tau(M_0) \sim \bigl( M_0^d l_{\rm P}^{2(d-1)} \bigr)^{\frac{1}{d-2}}.
\label{eq:BH_tau}
\end{equation}
In order for our minimal scheme to be able to describe the full formation 
and evaporation history of this black hole, $\tau(M_0)$ must be smaller 
than order $R$, yielding the condition for the Schwarzschild radius 
of the initial black hole $R_0 = [16\pi M_0 l_{\rm P}^{d-1}/(d-1) 
\Omega_{d-1}]^{1/(d-2)}$ as
\begin{equation}
  R_0 \lesssim \bigl( R l_{\rm P}^{d-1} \bigr)^{\frac{1}{d}}.
\label{eq:BH_cond}
\end{equation}
For AdS$_5 \times X_5$, this gives $R_0 \lesssim R^{1/4} l_{\rm P}^{3/4}$, 
which is inconsistent with the bound on $R_{\rm c}$ in Eq.~(\ref{eq:R_c}), 
$R_{\rm c} \gtrsim R^{2/5} l_{\rm P}^{3/5}$, obtained under the assumption 
that $X_5$ has a single length scale.

This restriction does not affect the description of the Hawking 
emission process in flat-space quantum gravity using the scheme of 
Section~\ref{subsec:R_cutoff}---we can still form a black hole whose 
Schwarzschild radius is larger than the lower bound on $R_{\rm c}$ and 
study how it emits Hawking quanta.  It does, however, prevent us from 
obtaining the full $S$-matrix elements between the initial collapsing 
matter and final Hawking quanta within the minimal scheme.  To obtain 
the $S$-matrix elements, one would need to modify the minimal scheme. 
This can be done, for example, by introducing absorptive boundary 
conditions at $r \sim R$ for $\tau > t_0$ by adiabatically turning 
on couplings between the dual CFT to another larger theory at 
$\tau \sim t_0$ through appropriate CFT operators (i.e.\ by introducing 
such couplings in the appropriate region in the $d$-dimensional 
Euclidean spacetime).  This would allow us to study the full evolution 
of a $(d+1)$-dimensional black hole, except for the last moment of 
evaporation where the physics is $(d+n+1)$-dimensional.

Our main interest in the rest of this paper is not to obtain the full 
$S$-matrix elements, so we will not be concerned with this issue. 
(Furthermore, it seems possible to circumvent the problem.)  We take 
the viewpoint that the scheme described thus far indeed works so that 
we can study evaporating flat-space black holes in $d+1$ dimensions 
using AdS/CFT duality.

\subsection{Time evolution}
\label{subsec:evol}

How can we describe time evolution in our (IR cut-off) flat spacetime? 
Recall that the set of CFT operators ${\cal B}$ in Eq.~(\ref{eq:set-B}) 
provides a complete basis for describing a single AdS volume at the 
center at $\tau = t_0$.  More precisely, a state representing a physical 
configuration in the region within the codimension-2 hypersurface 
$\partial V$ defined at $\tau = t_0$ can be written in general as
\begin{equation}
  \ket{f} = \sum_A f_A \Psi_A \ket{0},
\label{eq:f}
\end{equation}
where $\sum_A |f_A|^2 = 1$.  Now, similarly to $\Psi_A$'s, we can select 
CFT operators $\tilde{\Psi}_A$ representing physical configurations in 
the $d$-dimensional volume $\tilde{V}$
\begin{equation}
  0 \leq r < R, \qquad \tau = t_0 + t.
\label{eq:V-tilde}
\end{equation}
Here, as in the case of $V$, the time $\tau$ is specified at the boundary 
of the region $\partial \tilde{V}$ in the presence of arbitrary excitations.

Because of the symmetry, we can choose an operator $\tilde{\Psi}_A$ 
to represent the physical configuration on $\tilde{V}$ obtained by 
time-translating the configuration on $V$ created by $\Psi_A$ (with 
the same value of $A$) by the amount $\varDelta \tau = t$.  In terms 
of the coefficients $\alpha_{A,I}$ in Eq.~(\ref{eq:Psi_A}), this gives
\begin{equation}
  \tilde{\Psi}_A = \sum_{I \in \{ I| \Delta_I < c \} }\!\! 
    \alpha_{A,I}\, e^{i \Delta_I \frac{t}{R}} {\cal O}_I.
\label{eq:t-Psi_A}
\end{equation}
We call the set of operators obtained in this way $\tilde{\cal B}$:
\begin{equation}
  \tilde{\cal B} = \{ \tilde{\Psi}_A \}.
\label{eq:set-tilde-B}
\end{equation}
The number of elements of this set, $N_{\tilde{\cal B}}$, is obviously 
the same as that of ${\cal B}$, so that $\ln N_{\tilde{\cal B}} \sim c$; 
see Eq.~(\ref{eq:holo-CFT}).  Note that while the choice of $t_0$ is 
a convention, that of $t$ is not---it affects relations between $\Psi_A$'s 
and $\tilde{\Psi}_A$'s.

With this machinery, we can represent an arbitrary physical configuration 
on a single AdS volume by a set of coefficients $f_A$ with $\sum_A 
|f_A|^2 = 1$, where $A = 1, \cdots, N_{\cal B}$.  The amplitude for 
a configuration represented by $\{ f_A \}$ to become that represented 
by $\{ g_A \}$ after time $t$ is then given by
\begin{equation}
  \inner{g}{f} = \sum_{A,B=1}^{N_{\cal B}} g_A^* f_B\, 
    \bra{0} \tilde{\Psi}_A^\dagger \Psi_B \ket{0}.
\label{eq:amplitude}
\end{equation}
In terms of the coefficients $\alpha_{A,I}$, it can be written as
\begin{equation}
  \inner{g}{f} = \sum_{A,B}\, g_A^* \Gamma_{AB} f_B,
\label{eq:amplitude-2}
\end{equation}
where
\begin{equation}
  \Gamma_{AB} = \sum_{I \in \{ I| \Delta_I < c \} }\!\! 
    \alpha_{A,I}^*\, e^{-i \Delta_I \frac{t}{R}} \alpha_{B,I}.
\label{eq:Gamma_AB}
\end{equation}
Note that the (implicit) dependence of $\alpha_{A,I}$ on $t_0$ drops 
out from the amplitudes.

Since the CFT description is supposed to eliminate all the gauge 
redundancies in the gravitational theory (including those associated 
with the holographic reduction of the degrees of freedom), this provides 
a fully gauge-fixed description of physical observables in quantum 
gravity, i.e.\ correlations between physical entities (such as causal 
relations among events).  The time $t$ here corresponds to that measured 
at the asymptotic boundary of flat space.%
\footnote{This statement becomes formally exact (only) in the limit 
 $\epsilon \rightarrow 0$.}
Note that the matrix $\Gamma_{AB}$ in Eq.~(\ref{eq:Gamma_AB}) is 
not unitary because of the possibility that excitations come into 
or go outside the region under consideration between $\tau = t_0$ 
and $t_0 + t$.  Namely, ${\cal B}$ and $\tilde{\cal B}$ do not represent 
exactly the same set of CFT operators.  This effect, however, is 
negligible for processes occurring in a sufficiently inner region 
of the AdS volume (which corresponds to considering only certain 
subsets of $\Psi_A$'s and $\tilde{\Psi}_A$'s) within a time scale 
sufficiently shorter than the AdS scale (corresponding to taking $t$ 
sufficiently smaller than $R$).  Indeed, these are the processes that 
concern us below.

We mention that possible nonlinear instability in AdS spacetime is not 
an issue here.  Since theories in AdS with reflective boundary conditions, 
which we assume (at least) for $\tau < t_0$, are unitary, we can prepare 
any initial state at $t_0$, although it may correspond to finely-tuned 
quantum configurations at earlier times.  The freedoms in choosing $t_0$ 
and the conformal compactification of ${\bf S}^{d-1} \times {\bf R}$ to 
Euclidean ${\bf R}^d$ (specifically the relative scale between $t$ and 
$|x|$) correspond to the ambiguity in relating the initial states to 
the operators at $\tau = -\infty$.  This ambiguity, however, completely 
disappears from the transition amplitudes in Eqs.~(\ref{eq:amplitude-2},%
~\ref{eq:Gamma_AB}).

\section{Microscopic Dynamics of Spacetime in AdS/CFT}
\label{sec:BH-AdS}

In this section, we discuss what the physics of flat-space black holes 
implies for CFTs.  As has been emphasized, we focus our attention on 
the class of CFTs that admit weakly coupled gravitational descriptions 
in ($d+1$)-dimensional spacetime, although we expect that much of 
our analysis applies (with appropriate modifications) to the case 
in which the gravitational bulk has higher dimensions already at 
the AdS length scale.

At the level of perturbative expansion in $1/N$, where $N$ is 
related with $c$ by $c \sim N^2$, conditions for CFTs to have 
relevant gravitational descriptions were studied, e.g., in 
Refs.~\cite{Heemskerk:2009pn,Fitzpatrick:2010zm,ElShowk:2011ag,Kabat:2011rz}. 
In particular, it was argued that important ingredients are
\begin{itemize}
\item
large central charge, $c \gg 1$;
\item
the existence of a low-lying sector of operators whose correlators 
almost factorize, in particular a sector that admits $1/N$-type expansion;
\item
a large gap in the spectrum of operator dimensions, making string 
states heavy.
\end{itemize}
If these conditions are indeed sufficient, as demonstrated in simple 
cases in Ref.~\cite{Heemskerk:2009pn}, then the universal nature of black 
hole physics implies that our analysis can be viewed as predictions for 
nonperturbative properties of CFTs characterized by these ingredients.

Below, we discuss what the picture of evaporating black holes proposed 
in Refs.~\cite{Nomura:2014woa,Nomura:2014voa} to avoid the information 
problem~\cite{Hawking:1976ra,Almheiri:2012rt} implies for CFTs in the 
framework described in the previous section.  Since CFTs are well-defined, 
this in principle allows us to test (aspects of) the proposal, perhaps 
with the future development of techniques to analyze CFTs.

\subsection{General considerations}
\label{subsec:general}

What does the existence of a weakly coupled gravitational description 
in spacetime mean in our formalism?  As discussed in the previous section, 
we describe time evolution in quantum gravity in a fully gauge-fixed 
manner, through Eqs.~(\ref{eq:amplitude-2},~\ref{eq:Gamma_AB}).  In 
particular, the generator of time evolution is given by the Hamiltonian
\begin{equation}
  H_{AB} = \sum_{I \in \{ I| \Delta_I < c \} }\!\! 
    \alpha_{A,I}^* \frac{\Delta_I}{R} \alpha_{B,I}.
\label{eq:H_AB}
\end{equation}
In this language, the existence of a local gravitational description 
means the existence of an operator basis $\{ \Psi_A = \sum_I \alpha_{A,I} 
{\cal O}_I \}$ in which the action of $H_{AB}$ on some subsector(s) 
takes a local (nearest-neighbor) form.  (For a related discussion, see 
Ref.~\cite{Nomura:2011rb}.)  We will make this statement more precise 
in the next subsection, but for now this schematic picture is sufficient.

An interesting feature of Eq.~(\ref{eq:H_AB}) (or Eqs.~(\ref{eq:amplitude-2},%
~\ref{eq:Gamma_AB})) is that its structure is fully determined by the 
spectrum of operators, i.e.\ their dimensions, $\Delta_I$, and spins. 
On the other hand, we know that a CFT is specified by the spectrum of 
primary operators ${\cal O}_{\tilde{I}}$ as well as a set of constants 
$C^{\tilde{K}}_{\tilde{I}\tilde{J}}$ that appear in the OPE between 
primary fields (which are constructed by taking appropriate superpositions 
of primary and descendant operators).  What is the role played by these 
OPE coefficients, $C^{\tilde{K}}_{\tilde{I}\tilde{J}}$, in our formalism?

Let us first recall that a state $\Psi_A \ket{0}$ created by a basis 
operator represents a Heisenberg state of the ``entire universe'' in 
the gravitational picture.  Suppose we want to consider a process in 
which two high energy elementary particles (which are well localized 
in momentum space) collide to form a black hole, which then evaporates. 
This entire process then corresponds to some operator $\Psi \equiv \sum_A 
f_A \Psi_A = \sum_I \beta_I {\cal O}_I$, where the coefficients $\beta_I 
= \sum_A f_A \alpha_{A,I}$ have significant support only for values of 
$I$ in which the corresponding operators ${\cal O}_I$ have dimensions 
in a narrow range around some $\bar{\Delta} > c^{1/(d-1)}$ (reflecting 
the fact that the process occurs in a trans-Planckian regime).%
\footnote{If the black hole formed by the collision is large, with the 
 lifetime of order $R$ or larger, then we need to modify our minimal 
 scheme to describe the entire evaporation process, as discussed 
 in Section~\ref{subsec:dual}.  This issue is not relevant for our 
 discussion here.}
Note that all the relevant $\Psi_A$'s (or ${\cal O}_I$'s) here simply 
correspond to states with trans-Planckian energies.  How can we then 
interpret that the state represented by $\Psi$ consisted of two elementary 
particles at some time before the formation of the black hole?  More 
generally, how can we decompose a state (or the Hilbert space) of the 
entire universe into those of smaller subsystems?  This is where the 
OPE comes into play.

In our description, the OPE coefficients appear in the definition of 
the action of operators ${\cal O}_I$ on general (not necessarily vacuum) 
states.  For example, if a primary operator ${\cal O}_{\tilde{I}}$ acts 
on a state $\ket{\tilde{J}} = {\cal O}_{\tilde{J}} \ket{0}$ created by 
a primary operator ${\cal O}_{\tilde{J}}$, then the resulting state can 
be expanded as
\begin{equation}
  {\cal O}_{\tilde{I}} \ket{\tilde{J}} 
  \propto \Biggl( \sum_{\tilde{K}} C^{\tilde{K}}_{\tilde{I}\tilde{J}} 
    \ket{\tilde{K}} + \cdots \Biggr),
\label{eq:OPE}
\end{equation}
where the dots represent contributions from states corresponding to 
descendant operators, whose coefficients are determined by the conformal 
symmetry.  The action of an arbitrary operator ${\cal O}_I$ (primary 
or descendant) on an arbitrary state $\ket{J} = {\cal O}_{J} \ket{0}$ 
follows from conformal symmetry.

Now, let us consider the operator in the above example
\begin{equation}
  \Psi = \sum_I \beta_I {\cal O}_I,
\label{eq:Psi_BH}
\end{equation}
corresponding to a state that represents a black hole existing at 
$\tau = t_0$.  The statement that this black hole formed from two 
high energy elementary particles at some time $t_i$ ($< t_0$) is then 
translated into the statement that when the operator
\begin{equation}
  \tilde{\Psi} = \sum_I \beta_I\, 
    e^{i \Delta_I \frac{\delta t}{R}} {\cal O}_I,
\quad
  \delta t > t_0 - t_i,
\label{eq:tilde-Psi_BH}
\end{equation}
is expanded in terms of the basis operators $\Psi_A$ defined at 
$\tau = t_0$ as
\begin{equation}
  \tilde{\Psi} = \sum_A \tilde{g}_A \Psi_A,
\label{eq:tilde-Psi_exp}
\end{equation}
then the coefficients $\tilde{g}_A$ have significant support only 
for terms in which $\Psi_A$ are obtained by the OPEs of two operators 
representing single-particle states at $\tau = t_0$ (in the sense given 
in Eq.~(\ref{eq:OPE})).  We can write this schematically in the form
\begin{equation}
  \tilde{\Psi} \sim \Phi_1 \Phi_2,
\label{eq:2-particles}
\end{equation}
where $\Phi_1$ and $\Phi_2$ are single-particle operators (superposed 
appropriately to form wavepackets) which correspond to states having 
a single particle at $\tau = t_0$.  Note that single-particle operators 
can be defined purely algebraically at the nonperturbative level.  They 
are superpositions of components of low-lying conformal multiplets so 
that their OPE property and responses to the dilatation allow them to 
be viewed as the generators of a Fock-space like structure.  (In the 
language of CFT fields, this corresponds to the approximate factorization 
property of correlators.)

We stress that if we perform a similar analysis on $\Psi$, instead of 
the time translated $\tilde{\Psi}$, then the same statement need not apply. 
Specifically, if we expand $\Psi$ in terms of the basis operators $\Psi_A$
\begin{equation}
  \Psi = \sum_A g_A \Psi_A,
\label{eq:Psi_exp}
\end{equation}
then operators contributing to the right-hand side need not be simple 
multi-particle operators.  In fact, we assert that they are not.  Namely, 
operators appearing in Eq.~(\ref{eq:Psi_exp}) cannot be obtained by 
successive OPEs of single-particle operators within the regime in which 
an approximate particle interpretation holds.  In the gravitational 
description, this corresponds to the fact that states representing black 
holes cannot be obtained by a simple Fock space construction built on 
a flat spacetime background---they must be viewed as classical backgrounds 
on which the concept of particles is defined.

The existence of operators such as $\Psi$ here, which cannot be interpreted 
as multi-particle operators without shifting in time, has profound 
implications for how the semiclassical picture emerges from the fundamental 
theory of quantum gravity.  We now turn to this issue.

\subsection{Emergence of the semiclassical picture}
\label{subsec:semiclassical}

Recall that our basis operators, $\Psi_A$, are in one-to-one correspondence 
with possible physical configurations within our IR cut-off flat space, 
$V$, at some time $\tau = t_0$.  The number of such operators is given by 
Eq.~(\ref{eq:holo-CFT}), i.e.
\begin{equation}
  A = 1, \cdots, \exp\bigl[ O(c) \bigr].
\label{eq:A-range}
\end{equation}
In a CFT with a weakly coupled gravitational description, there are 
a class of operators $\Psi_{A_0}$ ($\subset \Psi_A$) that correspond 
to multi-particle states in a pure AdS background at $\tau = t_0$.  These 
operators obey a $1/N$-type scaling and can be identified by their OPE 
properties~\cite{Heemskerk:2009pn,Fitzpatrick:2010zm,ElShowk:2011ag,%
Kabat:2011rz}.  How many such operators exist nonperturbatively, and 
what is the distribution of their scaling dimensions?

To address these questions, we consider the bulk picture.  The highest 
entropy states consisting of multiple particles within the single AdS 
volume $V$ correspond to a thermal state.  Let the temperature of this 
system be $T$.  Then, its energy $E$ and (coarse-grained) entropy $S$ 
are given by
\begin{equation}
  E \sim T^{d+1} R^d,
\qquad
  S \sim T^d R^d.
\label{eq:E-S}
\end{equation}
The condition that the backreaction to spacetime is negligible (or 
sufficiently small) is given by the requirement that the Schwarzschild 
radius of a black hole of mass $E$ is smaller than $R$:
\begin{equation}
  (l_{\rm P}^{d-1} E)^{\frac{1}{d-2}} \lesssim R.
\label{eq:no-BH}
\end{equation}
Combining Eqs.~(\ref{eq:E-S}) and (\ref{eq:no-BH}), we obtain
\begin{equation}
  S \lesssim \left( \frac{R}{l_{\rm P}} \right)^{\frac{d(d-1)}{d+1}} 
    \sim O\bigl(c^{\frac{d}{d+1}}\bigr).
\label{eq:S-bound}
\end{equation}
This implies that the index $A_0$ runs only over an extremely small 
subset of the whole $A$ index
\begin{equation}
  A_0 = 1, \cdots, \exp\Bigl[ O\bigl(c^{\frac{d}{d+1}}\bigr) \Bigr].
\label{eq:A0-range}
\end{equation}
Namely, the degrees of freedom described by a semiclassical field theory 
on a fixed background comprise only a tiny subset of the whole quantum 
gravitational degrees of freedom~\cite{'tHooft:1993gx,Nomura:2013lia}. 

The distribution of the dimensions of operators $\Psi_{A_0}$ can be 
obtained from Eq.~(\ref{eq:E-S}).  We find that the number of $\Psi_{A_0}$ 
operators with dimensions between $\Delta$ and $\Delta + d\Delta$ is 
given by
\begin{equation}
  dN_{\Psi_{A_0}} \sim 
    \exp\Bigl[ O\bigl(\Delta^{\frac{d}{d+1}}\bigr) \Bigr]\, d\Delta,
\label{eq:Psi_A0-distr}
\end{equation}
in the relevant range of $0 < \Delta \lesssim c$.  For simplicity, here 
we have assumed that all the extra dimensions are small.  If some of them 
are large, so that the temperature of the multi-particle system exceeds 
the compactification scale $1/R_{\rm c}$, then the expression must 
be modified accordingly.

What do the rest of the operators, $\Psi_A$ with $A \notin \{ A_0 \}$, 
represent?  We claim that these operators correspond to states with 
nontrivial spacetime backgrounds at $\tau = t_0$, like $\Psi$ in the 
previous subsection.  In particular, they represent predominantly states 
with a black hole(s) at time $\tau = t_0$.%
\footnote{In this paper, we focus on black holes that are well approximated 
 by the Schwarzschild black hole.  We do not expect difficulty in 
 extending the analysis to more general cases.  For some implications 
 of charged black holes at the nonperturbative level in AdS/CFT, see 
 Ref.~\cite{Nakayama:2015hga}.}
The (coarse-grained) entropy of these states with total energy $E$ is 
given by the entropy of a single black hole of mass $E$ (if it is large 
enough) or the entropy of a thermal gas of total energy $E$ (if black 
holes compose only a small part of the entire system).  We thus find 
that the dimensions of operators $\Psi_A$ ($A \notin \{ A_0 \}$) is 
distributed as
\begin{equation}
  dN_{\Psi_{A\, \notin \{ A_0 \}}} \sim \left\{ \begin{array}{ll} 
    \exp\Biggl[ O\biggl( \frac{\Delta^{\frac{d-1}{d-2}}}{c^{\frac{1}{d-2}}} 
      \biggr) \Biggr]\, d\Delta 
      & \mbox{for } c^{\frac{d+1}{2d-1}} \lesssim \Delta \lesssim c,
\vspace{2mm} \\
    \exp\Bigl[ O\bigl(\Delta^{\frac{d}{d+1}}\bigr) \Bigr]\, d\Delta 
      & \mbox{for } c^{\frac{1}{d-1}} \lesssim \Delta 
      \lesssim c^{\frac{d+1}{2d-1}}.
    \end{array} \right.
\label{eq:Psi_A-distr}
\end{equation}
Here, we have again assumed that all the extra dimensions are small, of 
order $l_{\rm P}$ ($\sim l_{\rm s}$).  Note that for a finite value of 
$c$, a basis operator $\Psi_{A_0}$ cannot in general be chosen as one 
of the primary or descendant operators ${\cal O}_I$:\ $[D, \Psi_{A_0}] 
\,\slashed{\propto}\, \Psi_{A_0}$.  This implies that if it is time 
translated, it becomes an operator containing $\Psi_A$'s with $A \notin 
\{ A_0 \}$.  Therefore, some (not all) of the operators $\Psi_A$ with 
$A \notin \{ A_0 \}$ can be produced from $\Psi_{A_0}$'s through time 
evolution.

The existence of a black hole in the system does not mean that there 
are no degrees of freedom described by local dynamics.  We can certainly 
consider semiclassical field theory on a black hole background, which 
we believe can describe the dynamics of (at least) some degrees of freedom, 
even in the vicinity of the black hole.  How exactly do we construct 
a semiclassical (field or string) theory from the viewpoint of the 
fundamental theory of quantum gravity?

To illustrate the basic idea in the simplest manner, let us take the limit 
in which all the extra dimensions are small, as in Eq.~(\ref{eq:2-scales}). 
(Incorporating the effect of larger extra dimensions is straightforward.) 
The first step is to divide the basis operators $\Psi_{A\, \notin \{ A_0 \}}$ 
into groups $\Psi_{A_1}, \Psi_{A_2}, \cdots$ in such a way that members 
of each group represent possible states inside $\partial V$ that all 
correspond to (a time slice of) a specific spacetime geometry.  For 
example, we may take $\Psi_{A_1}$'s to represent states that have a 
black hole of mass $M$ (within the uncertainty of order the Hawking 
temperature $T_{\rm H}(M)$) at a specific location (within a proper 
length uncertainty of order $l_{\rm P}$).  Note that the specification 
of a geometry must necessarily involve uncertainties arising from quantum 
mechanics~\cite{Nomura:2014woa,Nomura:2014voa,Nomura:2014yka}.  Including 
the case of a pure AdS background, we may divide the index $A$ as
\begin{equation}
  \{ A \} = \bigcup\limits_{i=0,1,2,\cdots} \{ A_i \}.
\label{eq:index-decomp}
\end{equation}
Here, we take $\{ A_i \} \cap \{ A_j \} = \emptyset$ for $i \neq j$. 
As long as we are interested in a sufficiently short time scale, the 
description of the evolution of the system (using Eqs.~(\ref{eq:amplitude-2},%
~\ref{eq:Gamma_AB})) does not require operators beyond those in a 
single group.  For instance, in the above example of $\{ \Psi_{A_1} \}$, 
representing a black hole of mass $M$, the system can stay within the 
regime described by $\Psi_{A_1}$'s for time scale of $\varDelta \tau 
\lesssim 1/T_{\rm H}(M)$.  We can therefore build a semiclassical theory 
applicable for a limited time scale, using only operators in the single 
group.  A semiclassical theory on a time-dependent background can then 
be constructed by ``patching'' theories obtained in this way, each of 
which describes (semiclassical) physics in a certain time interval.

How can we construct a ``component'' semiclassical theory---a semiclassical 
theory applicable to a limited time interval---from operators $\Psi_{A_i}$ 
in a single group?  To be specific, let us consider the case in which 
$\{ \Psi_{A_i} \}$ ($i \neq 0$) represents states with a black hole of 
mass $M$ at a specific location (within appropriate uncertainties) at 
$\tau = t_0$.  In the region far from the black hole, physics is well 
described by a semiclassical theory on a pure AdS background.  This 
implies that $\Psi_{A_i}$'s can be OPE-decomposed (approximately) as
\begin{equation}
  \Psi_{A_i} \sim \Psi_{a^{\rm far}} \Psi_{A^{\rm near}},
\label{eq:Psi_Ai-decomp}
\end{equation}
where $\{ A_i \} = \{ a^{\rm far} \} \times \{ A^{\rm near} \}$, and 
$\Psi_{a^{\rm far}}$'s can be written as appropriate superpositions 
of $\Psi_{A_0}$'s corresponding to configurations of particles outside 
the near black hole region (often called the zone), conveniently defined 
to be the interior of the gravitational potential barrier for all the 
angular momentum modes:
\begin{equation}
  \tilde{r} \lesssim \biggl[ \frac{8\pi d}{(d-1) \Omega_{d-1}} 
    \biggr]^{\frac{1}{d-2}} (M l_{\rm P}^{d-1})^{\frac{1}{d-2}} 
  \equiv \tilde{r}_{\rm z},
\label{eq:zone}
\end{equation}
where $\tilde{r}$ is the Schwarzschild radial coordinate.  The black 
hole physics is encoded in the structure of $\{ \Psi_{A^{\rm near}} \}$ 
as well as the ``interactions'' between $\{ \Psi_{a^{\rm far}} \}$ and 
$\{ \Psi_{A^{\rm near}} \}$ (i.e.\ how $\Psi_{A_i}$'s are expanded in 
terms of $\Psi_{a^{\rm far}}$'s and $\Psi_{A^{\rm near}}$'s defined 
analogously for a black hole of a different mass, $\Psi_{A_j}$'s with 
$j \neq i$, after shifting in time by $\varDelta \tau \gtrsim 
1/T_{\rm H}(M)$).

The number of operators $\Psi_{A^{\rm near}}$ is, at least, of the 
order of the exponential of the Bekenstein-Hawking entropy of a black 
hole of mass $M$
\begin{equation}
  S_{\rm BH}(M) = \biggl[ \frac{4^d \pi^{d-1}}{(d-1)^{d-1} 
    \Omega_{d-1}} \biggr]^{\frac{1}{d-2}} (M l_{\rm P})^{\frac{d-1}{d-2}}.
\label{eq:S_BH}
\end{equation}
On the other hand, an analysis similar to the one that led to 
Eq.~(\ref{eq:A0-range}) implies that the number of possible configurations 
of semiclassical degrees of freedom within the zone, Eq.~(\ref{eq:zone}), 
is much smaller than that implied by this number; specifically, it is 
given by the exponential of
\begin{equation}
  S_{\rm sc}(M) \sim S_{\rm BH}(M)^{\frac{d}{d+1}} \ll S_{\rm BH}(M).
\label{eq:S_sc}
\end{equation}
Here, by the semiclassical degrees of freedom we mean the degrees of 
freedom that have ``reference frame independent'' meaning within the 
semiclassical theory, e.g.\ a detector located in the zone.  In particular, 
they do not include the degrees of freedom associated with the ``thermal 
atmosphere'' of the black hole, whose existence is frame dependent.  (This 
point will be discussed in more detail in the following subsections.) 
This suggests that we can label $\Psi_{A^{\rm near}}$'s using two indices 
$a$ and $k$, each of which refers to the (reference frame independent) 
semiclassical degrees of freedom in the zone and the rest of the degrees 
of freedom represented by the $\Psi_{A^{\rm near}}$'s.  The latter are 
called the vacuum degrees of freedom~\cite{Nomura:2014woa,Nomura:2014voa}.

The analysis described above indicates that the index $a$ runs over
\begin{equation}
  a = 1, \cdots, \exp\bigl[ S_{\rm sc}(M) \bigr].
\label{eq:a-range}
\end{equation}
In general, the physics of the vacuum degrees of freedom, $k$, does 
not decouple from that of the semiclassical degrees of freedom, $a$ 
and $a^{\rm far}$.  This effect, however, is negligible for sufficiently 
short time scales in which the information exchange between these degrees 
of freedom can be ignored---in our context, time scales shorter than 
the single Hawking emission time of order $1/T_{\rm H}(M)$.  In this 
case, we can take $k$ to run over a fixed range
\begin{equation}
  k = 1, \cdots, \exp\bigl[ S_{\rm BH}(M) \bigr].
\label{eq:k-range}
\end{equation}
A basis operator in the group, $\in \{ \Psi_{A_i} \}$, can then be 
specified by the set of indices
\begin{equation}
  \{ A_i \} = \{ a^{\rm far} \} \times \{ (a, k) \},
\label{eq:A_i}
\end{equation}
where $a$ and $k$ run over the ranges given by Eqs.~(\ref{eq:a-range},%
~\ref{eq:k-range}).%
\footnote{To be more precise, we also need an index $\bar{a}$ labeling 
 the modes on the stretched horizon, which is located at a microscopic 
 distance outside of the mathematical horizon and is regarded as a physical 
 (timelike) membrane that may be physically excited~\cite{Susskind:1993if}. 
 We will omit this index below because it is not essential for our 
 discussion here, but including it is straightforward.  (For example, 
 Eq.~(\ref{eq:A_i}) becomes $\{ A_i \} = \{ a^{\rm far} \} \times 
 \{ (a, \bar{a}, k) \}$.)  For more details about these modes, see 
 Refs.~\cite{Nomura:2014woa,Nomura:2014voa}.}
We emphasize that we focus here only on a component semiclassical theory 
applicable in time scales sufficiently shorter than $\sim 1/T_{\rm H}(M)$. 
The non-decoupling nature of the semiclassical and vacuum degrees of 
freedom becomes important for physics of longer time scales, especially 
the evolution of the black hole discussed in the next subsection. 
In the language adopted here, this has to do with how we patch different 
component theories together to obtain a full semiclassical theory on 
a time-dependent background.

The component semiclassical theory in question is obtained by coarse-graining 
the vacuum degrees of freedom, represented by $k$.  By assumption, 
interactions between the semiclassical and vacuum degrees of freedom 
are negligible, implying that the generator of time evolution at the 
microscopic level---$H_{(a^{\rm far},a,k) (a^{{\rm far}\prime},a',k')}$ 
given by Eq.~(\ref{eq:H_AB})---factor as
\begin{equation}
  H_{(a^{\rm far},a,k) (a^{{\rm far}\prime},a',k')} 
  \sim H^{({\rm sc})}_{(a^{\rm far},a) (a^{{\rm far}\prime},a')} 
    H^{({\rm vac})}_{k k'}.
\label{eq:H-factor}
\end{equation}
The state of the semiclassical theory representing the configuration 
$(a^{\rm far},a)$ can be identified as the maximally mixed state in 
$k$ space:
\begin{equation}
  \rho_{a^{\rm far}, a} = \frac{1}{e^{S_{\rm BH}}} 
    \sum_{k=1}^{e^{S_{\rm BH}}} \ket{\Psi_{a^{\rm far},a,k}} 
    \bra{\Psi_{a^{\rm far},a,k}},
\label{eq:rho_a}
\end{equation}
where $\ket{\Psi_{a^{\rm far},a,k}} \equiv \Psi_{a^{\rm far},a,k} \ket{0}$. 
Since physics in $k$ space do not concern us here, we may choose its 
basis to agree with the approximate energy eigenstates (within the 
time scale of interest), $H^{({\rm vac})}_{k k'} \sim \delta_{k k'}$. 
The microscopic time evolution, obtained by acting Eq.~(\ref{eq:H-factor}) 
on Eq.~(\ref{eq:rho_a}), can then be written as if $\rho_{a^{\rm far}, a}$ 
is a quantum field theory state evolving under the semiclassical (gauge-fixed) 
``Hamiltonian'' $H^{({\rm sc})}_{(a^{\rm far},a) (a^{{\rm far}\prime},a')}$:
\begin{equation}
  \rho_{a^{\rm far}, a} 
  \sim \bigl| \Psi^{({\rm sc})}_{a^{\rm far},a} \bigr\rangle
    \bigl\langle \Psi^{({\rm sc})}_{a^{\rm far},a} \bigr|.
\label{eq:Psi_sc}
\end{equation}
From the point of view of the fundamental theory, the 
semiclassical theory emerges because the labeling scheme in 
Eq.~(\ref{eq:A_i}) can be chosen such that the Hamiltonian 
$H^{({\rm sc})}_{(a^{\rm far},a) (a^{{\rm far}\prime},a')}$ gives 
time evolution of $\ket{\Psi^{({\rm sc})}_{a^{\rm far},a}}$ in such 
a way that it occurs on a fixed black hole background and is local 
at length scales larger than $l_{\rm s}$.%
\footnote{If we include the stretched horizon modes, the 
 relevant Hamiltonian takes the form $H_{(a^{\rm far},a,\bar{a}) 
 (a^{{\rm far}\prime},a',\bar{a}')}$.  The dynamics of the $\bar{a}$ 
 modes represented by this Hamiltonian need not be local in the 
 angular directions.}
Remember that our component semiclassical theory here is applicable 
only for time scales shorter than $\sim 1/T_{\rm H}(M)$.  Also, since 
our time variable is determined at the boundary of the space, we expect 
that it corresponds (approximately) to the Schwarzschild time.  The 
semiclassical picture obtained here, therefore, provides a distant 
view of the black hole.

\subsection{The evolution of a flat-space black hole}
\label{subsec:BH}

In order to discuss the evolution of a black hole, we need to consider 
the dynamics at time scales longer than $\sim 1/T_{\rm H}(M)$.  At the 
level of the fundamental theory, the evolution of the system is simply 
described by the Hamiltonian in Eq.~(\ref{eq:H_AB}), giving unitary 
evolution of the black hole (within a time scale of order $R$).  However, 
to have a simple framework to interpret the evolution process, we need to 
match to the semiclassical picture.  This can be done by patching together 
the component semiclassical theories described in the previous subsection.

Let us consider two groups of basis operators $\{ \Psi_{A_i} \}$ and 
$\{ \Psi_{A_j} \}$ each representing states with a black hole of mass 
$M_i$ and $M_j$ with $M_i - M_j \sim T_{\rm H}(M_i)$, located at a specific 
position.%
\footnote{We focus only on an evaporating black hole.  The $CPT$ 
 invariance of the theory implies that there are an equal number of 
 states involving the corresponding ``anti-evaporating'' black hole. 
 Since these states can only be formed from exponentially fine-tuned 
 initial states, we do not consider them.}
(Here appropriate uncertainties for both the mass and position are 
implied.  The precise meaning of the black hole mass in our context 
will be discussed shortly.)  We take the basis operators in two groups 
to be orthogonal, $\inner{\Psi_{A_i}}{\Psi_{A_j}} = 0$ for $i \neq j$, 
and we assume that these operators provide a complete basis to describe 
the evolution of a black hole in the relevant time scale of order 
$1/T_{\rm H}$; in other words, by preparing a series of groups similarly, 
i.e.\ $\{ \Psi_{A_i} \}, \{ \Psi_{A_j} \}, \{ \Psi_{A_k} \}, \cdots$ 
representing the black hole of mass $M_i, M_j = M_i - T_{\rm H}(M_i), 
M_k = M_j - T_{\rm H}(M_j)$, and so on, we can fully describe the future 
evolution of an initial black hole of mass $M_i$ (modulo macroscopic 
effects discussed in Ref.~\cite{Page:1979tc,Nomura:2012cx}, which 
can be included straightforwardly by adding groups representing black 
holes at different locations, spins, and charges).  We assume that the 
approximation of Eq.~(\ref{eq:H-factor}) holds for evolution within 
each group, and that nontrivial physics associated with Hawking 
emission---in particular, interactions between semiclassical and 
vacuum degrees of freedom---occurs only when the system evolves 
between the groups.  Of course, this ``discretization'' of the emission 
process is an approximation, and the true microscopic evolution occurs 
continuously through the Hamiltonian in Eq.~(\ref{eq:H_AB}).

We define the black hole mass $M$ to be the whole energy of the 
{\it evolving} black hole system (measured in the asymptotic region) 
except for the part associated with the semiclassical degrees of 
freedom that have reference frame independent meaning.  In particular, 
it contains energies associated with the thermal atmosphere of the 
black hole as well as an ingoing negative energy flux that appears 
in the calculation of the stress-energy tensor on the relevant 
background~\cite{Davies:1976ei}.  Therefore, our vacuum degrees 
of freedom $k$ contain the degrees of freedom associated with these 
entities.%
\footnote{This definition is different from the one adopted mainly 
 in Refs.~\cite{Nomura:2014woa,Nomura:2014voa}, although the physics 
 described below is equivalent.  Roughly speaking, the definition 
 used in Refs.~\cite{Nomura:2014woa,Nomura:2014voa} corresponds 
 to describing the system as an expansion around the Hartle-Hawking 
 vacuum~\cite{Hartle:1976tp} at each moment in time, while here we 
 describe it as an expansion around the evolving black hole background, 
 which is well approximated by the advanced/ingoing Vaidya spacetime 
 near the horizon~\cite{Bardeen:1981zz}.  See Section~3.2 of 
 Ref.~\cite{Nomura:2014voa} for a detailed discussion on this point.}
With this definition, the operators $\{ \Psi_{A_i} \}$ can be labeled 
by the indices $a^{\rm far}_i, a_i, k_i$ with $k_i$ running over
\begin{equation}
  k_i = 1, \cdots, \exp\bigl[ S_{\rm BH}(M_i) \bigr],
\label{eq:k_i-range}
\end{equation}
while $\{ \Psi_{A_j} \}$ by $a^{\rm far}_j, a_j, k_j$ with $k_j$ taking
\begin{equation}
  k_j = 1, \cdots, \exp\bigl[ S_{\rm BH}(M_j) \bigr].
\label{eq:k_j-range}
\end{equation}
Note that the latter range is smaller than the former because of the 
smaller mass of the black hole, $M_j < M_i$.  The Hawking emission, caused 
by $H_{A_j A_i}$ in Eq.~(\ref{eq:H_AB}), then occurs as
\begin{equation}
  \ket{\Psi_{a^{\rm far}_i,a_i,k_i}} 
  \rightarrow \sum_{a^{\rm far}_j,a_j,k_j}\!\! 
    \gamma_{a^{\rm far}_i,a_i,k_i;\, a^{\rm far}_j,a_j,k_j} 
    \ket{\Psi_{a^{\rm far}_j,a_j,k_j}},
\label{eq:Hawking-micro}
\end{equation}
in the time scale of order $1/T_{\rm H}(M_i)$.  In order for this process 
to be unitary, the space labeled by $a^{\rm far}_j$ must be larger 
than that by $a^{\rm far}_i$.  In fact, $a^{\rm far}_j$ contains 
the component labeling the newly emitted Hawking quanta, which was not 
present before the process occurred.%
\footnote{The sum in the right-hand side of Eq.~(\ref{eq:Hawking-micro}) 
 contains (small) components in which the energies of emitted Hawking 
 quanta are larger than the chosen uncertainties of $M_i$ and $M_j$. 
 In these components, $M_j$ should be understood to take different values 
 determined by the energy of the emitted Hawking quanta through energy 
 conservation.  Including this effect explicitly (which we will not do) 
 does not affect our discussion below. \label{ft:tail}}

An important aspect of the picture in Refs.~\cite{Nomura:2014woa,%
Nomura:2014voa} is that the process in Eq.~(\ref{eq:Hawking-micro}) 
must be viewed as occurring around the edge of the zone, $\tilde{r} 
\sim \tilde{r}_{\rm z}$.  In the present scheme, this manifests as 
follows.  $a^{\rm far}_i$ and $a_i$ (and $a^{\rm far}_j$ and $a_j$) 
label the largest possible degrees of freedom that can be interpreted 
as the semiclassical excitations obeying local dynamics on a fixed 
spacetime background.  To put it the other way around, the vacuum degrees 
of freedom, $k_i$ and $k_j$, do not admit such an interpretation.  The 
fact that a part of the information in $k_i$ is transferred directly 
to $a^{\rm far}_j$ in Eq.~(\ref{eq:Hawking-micro}) should then be 
interpreted from the semiclassical point of view that a part of the 
microscopic information about the black hole ($k_i$) is distributed at 
the edge of the zone, and that the information is transferred there to 
semiclassical degrees of freedom outside the zone ($a^{\rm far}_j$)---i.e.\ 
the Hawking quanta---without involving an information transportation 
mechanism from the horizon within the semiclassical theory.  Because 
of energy conservation, this process must be accompanied by the creation 
of an ingoing negative energy flux (reflected here by $M_j < M_i$), which 
carries {\it negative entropy}~\cite{Nomura:2014woa,Nomura:2014voa} 
(reflected by the fact that $k_j$'s run over smaller ranges than $k_i$).

The mechanism of information transfer described above avoids the 
paradox in Refs.~\cite{Almheiri:2012rt,Almheiri:2013hfa,Marolf:2013dba,%
Polchinski:2015cea} which relies crucially on the assumption that some 
information transportation mechanism is in operation from the horizon 
to the edge of the zone on a semiclassical background.  Our picture 
says that the information transfer from black holes cannot be understood 
in that manner.

Since the only essential ingredients in the process of 
Eq.~(\ref{eq:Hawking-micro}) are $k_i$, $k_j$, and (a part of) 
$a^{\rm far}_j$, we may suppress other indices and write it as
\begin{equation}
  \ket{\Psi_{k_i}} \rightarrow \sum_{h,k_j}\!\! 
    \gamma_{k_i;\, h,k_j} \ket{\Phi_h} \ket{\Psi_{k_j}},
\label{eq:Hawking-micro-2}
\end{equation}
where $h \subset a^{\rm far}_j$ labels the emitted Hawking quanta, 
and we have introduced the notation of separating the emitted quanta 
$\ket{\Psi_{h,k_j}} = \ket{\Phi_h} \ket{\Psi_{k_j}}$.  How does the 
semiclassical theory describe this process?  Given that the semiclassical 
theory is obtained after coarse-graining the vacuum degrees of 
freedom, Eqs.~(\ref{eq:rho_a},~\ref{eq:Psi_sc}), the description 
at the semiclassical level is that the initial state
\begin{equation}
  \rho_{\rm init} = \frac{1}{e^{S_{\rm BH}(M_i)}}\! 
    \sum_{k_i=1}^{e^{S_{\rm BH}(M_i)}} \ket{\Psi_{k_i}} \bra{\Psi_{k_i}},
\label{eq:sc_initial}
\end{equation}
evolves as
\begin{equation}
  \rho_{\rm init} \rightarrow 
    \frac{1}{e^{S_{\rm BH}(M_i)}}\! \sum_{k_i=1}^{e^{S_{\rm BH}(M_i)}} 
    \sum_{h,h'} \sum_{k_j,k'_j=1}^{e^{S_{\rm BH}(M_j)}} 
    \gamma_{k_i;\, h,k_j} \gamma^*_{k_i;\, h',k'_j} 
    \ket{\Phi_h} \ket{\Psi_{k_j}} \bra{\Phi_{h'}} \bra{\Psi_{k'_j}}.
\label{eq:sc_evol}
\end{equation}
Because of an enormous number of degrees of freedom involved, it is 
natural to expect that the sums over the vacuum degrees of freedom 
show the thermodynamic characteristics:
\begin{equation}
  \frac{1}{e^{S_{\rm BH}(M_i)}}\! \sum_{k_i=1}^{e^{S_{\rm BH}(M_i)}} 
    \sum_{k_j,k'_j=1}^{e^{S_{\rm BH}(M_j)}} 
    \gamma_{k_i;\, h,k_j} \gamma^*_{k_i;\, h',k'_j} 
    \ket{\Psi_{k_j}} \bra{\Psi_{k'_j}} 
  \approx \frac{1}{Z} g_h e^{-\frac{E_h}{T_{\rm H}(M_i)}} \delta_{hh'},
\label{eq:Hawking-stat}
\end{equation}
where $E_h$ is the energy of the state $\ket{\Phi_h}$, $Z = \sum_h g_h 
\exp[-E_h/T_{\rm H}(M_i)]$, and $g_h$ is a factor that depends on $h$. 
Then, Eq.~(\ref{eq:sc_evol}) becomes
\begin{equation}
  \rho_{\rm init} \rightarrow 
    \frac{1}{Z} \sum_h g_h e^{-\frac{E_h}{T_{\rm H}(M_i)}} 
    \ket{\Phi_h} \bra{\Phi_h}.
\label{eq:sc_evol-2}
\end{equation}

In contrast with the case in a component semiclassical theory, there is 
no way to write this evolution (obtained after coarse-graining the vacuum 
degrees of freedom) preserving unitarity within quantum field theory. 
The best one can do is to write it as
\begin{equation}
  \bigl| \Psi^{({\rm sc})}_{\rm vac}(M_i) \bigr\rangle
    \bigl\langle \Psi^{({\rm sc})}_{\rm vac}(M_i) \bigr| 
  \rightarrow \frac{1}{Z} \sum_h g_h e^{-\frac{E_h}{T_{\rm H}(M_i)}} 
    \bigl| \Psi^{({\rm sc})}_h(M_j) \bigr\rangle
    \bigl\langle \Psi^{({\rm sc})}_h(M_j) \bigr|.
\label{eq:Hawking-sc}
\end{equation}
This is Hawking's original result~\cite{Hawking:1974rv} with $g_h$ 
representing the gray-body factor calculable in the semiclassical 
theory~\cite{Page:1976df}.  We expect that modes softer than 
$T_{\rm H}(M)$ can still be described by the standard local dynamics 
in the patched semiclassical theory, except when they participate in 
the Hawking emission process.  In particular, we expect this to be the 
case even around the black hole, since sending generic soft quanta to 
an evaporating black hole is not the same as the time reversal of the 
Hawking emission process, which corresponds to sending highly fine-tuned 
quanta to an anti-evaporating black hole~\cite{Nomura:2014voa}.  (This 
is also consonant with the fact that Hawking emission is a process in 
which the coarse-grained entropy increases~\cite{Zurek:1982zz}.)

The above analysis explains why the semiclassical calculation of 
Refs.~\cite{Hawking:1974rv,Hawking:1976ra} finds an apparent violation 
of unitarity---semiclassical theory is incapable of resolving the 
microscopic dynamics of the vacuum degrees of freedom {\it by construction}, 
and hence is secretly dealing with mixed states~\cite{Nomura:2014woa,%
Nomura:2014voa,Nomura:2013lia,Nomura:2014yka}.  For sufficiently short 
time scales, this aspect can be neglected because interactions between 
the semiclassical and vacuum degrees of freedom are relatively slow. 
However, for longer time scales, i.e.\ when the Hawking emission effect 
becomes relevant, the mixed nature of the vacuum degrees of freedom is 
necessarily ``leaked'' into semiclassical degrees of freedom, allowing a 
description of Hawking quanta only as mixed states.  In our treatment here, 
this occurs because of the necessity of patching multiple component theories 
together, throughout which the approximation in Eq.~(\ref{eq:H-factor}) 
does not apply (although it holds within each of the component theories).

A similar analysis can be performed for black hole mining, in which 
the energy (and information) of a black hole is directly extracted 
by an apparatus located within the zone~\cite{Unruh:1982ic}.  The time 
scale for this process is still of order $1/T_{\rm H}(M)$ as measured 
in $\tau$~\cite{Brown:2012un}, although the number of available 
``channels'' is larger than that of spontaneous Hawking emission, 
so it can accelerate the energy loss rate of the black hole.  (This 
is because unlike the case of spontaneous Hawking emission, which 
is dominated by $s$-wave, higher angular momentum modes can also 
contribute to the mining process.)

The process can still be described as in Eq.~(\ref{eq:Hawking-micro}). 
There are essentially only two differences from the case of spontaneous 
Hawking emission.  First, the decrease of the range over which the vacuum 
degrees of freedom runs, Eqs.~(\ref{eq:k_i-range},~\ref{eq:k_j-range}), 
is compensated by the degrees of freedom representing excited states 
of the apparatus ($\subset a_j$), not by the Hawking quanta ($\subset 
a^{\rm far}_j$).  Second, since the process involves higher angular 
momentum modes, the negative energy excitations arising as backreactions 
can be localized in angular directions.  This implies that the final 
state in Eq.~(\ref{eq:Hawking-micro}) should now be interpreted as 
a black hole of mass $M_i$ with ingoing excitations of negative energy 
$M_j-M_i$ ($<0$) even within the semiclassical theory, as described 
in Ref.~\cite{Unruh:1983ms}.  (This implies that the labeling scheme 
in Eq.~(\ref{eq:A_i}) contains operators representing such excitations 
as a part of the modes labeled by $a$.)  Denoting the ground and excited 
states of the apparatus by $0$ and $x$ ($x = 1,2,\cdots$), respectively, 
and the negative energy excitations by $n$, we can write the equation 
analogous to Eq.~(\ref{eq:Hawking-micro-2}) as
\begin{equation}
  \ket{\Psi_{0,k_i}} \rightarrow \sum_{x,n,k_j}\!\! 
    \gamma_{k_i;\, x,n,k_j} \ket{\Psi_{x,n,k_j}}.
\label{eq:mining-micro}
\end{equation}
Note that the apparatus states and the negative energy excitations need 
not be maximally entangled after the process, which allows for transferring 
information from the black hole to the apparatus~\cite{Nomura:2014voa}.%
\footnote{For components in which $E_x$ is larger than the uncertainties 
 of $M_i$ and $M_j$, the extra energies needed to excited the apparatus 
 must be compensated by the negative energy excitations.  This creates 
 some amount of entanglement between the apparatus states and the 
 negative energy excitations.  (See also footnote~\ref{ft:tail} for 
 a related discussion.)  Below, we ignore this small effect.}
After coarse-graining the vacuum degrees of freedom and assuming 
the thermal property for these large degrees of freedom, as in 
Eq.~(\ref{eq:Hawking-stat}), we arrive at the equation analogous 
to Eq.~(\ref{eq:Hawking-sc}):
\begin{equation}
  \bigl| \Psi^{({\rm sc})}_{0}(M_i) \bigr\rangle
    \bigl\langle \Psi^{({\rm sc})}_{0}(M_i) \bigr| 
  \rightarrow \frac{1}{Z} \sum_x g_x e^{-\frac{E_x}{T_{{\rm H, loc}}(M_i)}} 
    \bigl| \Psi^{({\rm sc})}_{x,n}(M_i) \bigr\rangle
    \bigl\langle \Psi^{({\rm sc})}_{x,n}(M_i) \bigr|,
\label{eq:mining-sc}
\end{equation}
where $E_x$ are the proper energies needed to excite the apparatus 
from the ground to the $x$ states, and $T_{{\rm H, loc}}(M_i)$ is 
the local Hawking temperature at the location of the apparatus.  $g_x$ 
is the response function reflecting intrinsic properties of the apparatus 
under consideration, and $Z = \sum_x g_x \exp[-E_x/T_{{\rm H, loc}}(M_i)]$. 
$\ket{\Psi^{({\rm sc})}_{x,n}(M_i)}$ represents the semiclassical state 
in which, in addition to the apparatus in the $x$ state, the ingoing 
excitations with negative energy $M_j - M_i$ (as measured in the 
asymptotic region) are excited over the black hole of mass $M_i$.

Once again, in the semiclassical picture the process in 
Eq.~(\ref{eq:mining-micro}) occurs at the location of the apparatus 
without involving an information transportation mechanism from the 
horizon to there over a semiclassical spacetime background.  The 
negative energy excitations carry negative entropies as reflected 
by the fact that the ranges over which $k_j$'s run are smaller than 
that of $k_i$.  These negative energy excitations will scramble 
with the vacuum degrees of freedom in the time scale of order 
$(1/T_{\rm H}(M_i)) \ln(1/T_{\rm H}(M_i) l_{\rm P})$~\cite{Hayden:2007cs} 
after reaching the stretched horizon, making the system relax into 
a black hole of mass $M_j$ (other than the apparatus).

\subsection{Spacetime-matter duality}
\label{subsec:st-matter}

The CFT description of the black hole physics presented above sheds new 
light on the meaning of spacetime-matter duality~\cite{Nomura:2014woa,%
Nomura:2014voa}, the term referring to the fact that the black hole 
microstates can be viewed, in some sense, as playing roles of both 
spacetime and matter, but in fact are neither.%
\footnote{This is reminiscent of wave-particle duality in quantum 
 mechanics---a quantum object exhibits dual properties of waves 
 and particles while the ``true'' (quantum) description does not 
 fundamentally rely on either of these classical concepts.}

First, consistency with the semiclassical calculation---or the assumption 
of ``standard'' thermodynamic properties for the vacuum degrees of 
freedom---indicates that the black hole microstates, represented by 
$k$'s, interact with the semiclassical degrees of freedom as if they 
comprise a thermal atmosphere of the black hole (modulated by the negative 
energy flux due to the evolution) with the temperature given by the 
blue-shifted Hawking temperature
\begin{equation}
  T(\tilde{r}) = 
    \frac{1}{\sqrt{1-\frac{\tilde{r}_0^{d-2}}{\tilde{r}^{d-2}}}} 
    \frac{d-2}{4\pi \tilde{r}_0}\,\, \theta(\tilde{r}_{\rm z} - \tilde{r}),
\label{eq:T-thermal}
\end{equation}
where $\tilde{r}_0 = [16\pi M l_{\rm P}^{d-1}/(d-1) 
\Omega_{d-1}]^{1/(d-2)}$ is the Schwarzschild radius; see 
Eqs.~(\ref{eq:Hawking-sc},~\ref{eq:mining-sc}).  On the other hand, 
we may also interpret these microstates as the ``spacetime degrees 
of freedom'' arising from the fact that the uncertainty principle 
forces us to coarse-grain ``detailed'' spacetime geometry to arrive 
at the semiclassical picture:\ the number of independent quantum states 
representing black holes of mass between $M$ and $M + \varDelta M$ 
with $\varDelta M \approx O(T_{\rm H}(M))$ is labeled by the $k$ 
index~\cite{Nomura:2014yka}.  The vacuum degrees of freedom play dual 
roles of matter and spacetime~\cite{Nomura:2014woa,Nomura:2014voa}.

There is also a sense, however, in which the vacuum degrees of freedom 
cannot be usual matter or spacetime.  As already discussed in the 
previous subsection, the internal dynamics of these degrees of freedom 
cannot be organized in a way such that they are subject to local dynamics 
on some spacetime background.  In this sense, they are not real matter.%
\footnote{This has an interesting implication if the lesson from 
 black holes can be extrapolated to (meta-stable) de~Sitter spacetimes:\ 
 the problem of Boltzmann brains may be ``trivially'' solved because 
 the dynamics of the vacuum degrees of freedom may not support 
 intelligent observers~\cite{Nomura:2015zda}.}
Furthermore, these degrees of freedom also exhibit a peculiar feature 
called extreme relativeness~\cite{Nomura:2014woa,Nomura:2014voa}:\ 
while in the distant reference frame they can be viewed as distributed 
according to the thermal entropy calculated using Eq.~(\ref{eq:T-thermal}), 
in other reference frames their distribution is different.  (This 
feature cannot be seen in the analysis here which is tied to the distant 
description.)  Namely, these degrees of freedom are not ``anchored'' 
to spacetime, and in this sense they cannot be viewed as a spacetime 
itself, at least in the sense envisioned in standard general relativity.

What are the vacuum degrees of freedom then?  In the CFT description 
given here, they correspond to operators with high scaling dimensions, 
$\Delta \gtrsim c^{1/(d-1)}$, which cannot be interpreted as simple 
multi-particle states.  These operators are significant especially 
because they can be generated by simple multi-particle operators through 
time evolution.  Specifically, we can consider an operator $\Psi$ 
representing two energetic particles aimed at each other at some reference 
time $t_0$, which can be related (as in Eq.~(\ref{eq:2-particles})) to 
two single-particle operators by the OPE
\begin{equation}
  \Psi \sim \Phi_1 \Phi_2.
\label{eq:ex-init}
\end{equation}
Here, each $\Phi_i$ is an operator representing an appropriate 
single-particle wavepacket at time $t_0$, and is obtained by superposing 
(a tremendous number of) energy-eigenstate operators ${\cal O}_I$.  We 
can then time translate this operator by an amount $t$ using the dilatation 
operator $D$
\begin{equation}
  \Psi(t) \sim e^{-i D t}\, \Psi\, e^{i D t},
\label{eq:ex-evol}
\end{equation}
and ask how/if this operator can be expressed in terms of operators 
representing multi-particle states at $t_0$.  Recall that these multi-particle 
operators can be defined purely in a CFT at the nonperturbative level by 
their characteristic OPE structure.

Suppose the energy of the initial state is super-Planckian; namely, 
when $\Psi$ is expanded in terms of $D$-eigenstate operators, the 
coefficients are most significant for operators having dimensions 
larger than $c^{1/(d-1)}$.  If $t$ is sufficiently large, corresponding 
to a time longer than the lifetime of the black hole formed by the 
collision, then $\Psi(t)$ can be written as (a superposition of) 
multi-particle operators; schematically
\begin{equation}
  \Psi(t) \sim \sum_n \sum_{\Phi^{(n)}} a_{\Phi^{(n)}} \Phi^{(n)},
\label{eq:ex-fin}
\end{equation}
where $\Phi^{(n)}$ represents $n$-particle operators obtained from 
single-particle operators (operators representing single-particle states 
at $t_0$) by OPEs within the regime in which the approximate particle 
interpretation is possible.  This corresponds to a final Hawking 
radiation state.%
\footnote{Again, if the black hole formed by the collision is large 
 so that its lifetime exceeds $R$, then this minimal scheme needs to 
 be modified; see Section~\ref{subsec:dual}.  This does not affect 
 our discussion below.}
On the other hand, if $t$ is chosen to be smaller than the black hole's 
lifetime (but larger than the time it takes for the initial particles 
to collide), then $\Psi(t)$ cannot be approximated by multi-particle 
operators, i.e.
\begin{equation}
  \Psi(t) \nsim \sum_n \sum_{\Phi^{(n)}} a_{\Phi^{(n)}} \Phi^{(n)},
\label{eq:ex-BH-not}
\end{equation}
for any choice of $a_{\Phi^{(n)}}$.  Rather, $\Psi(t)$ is written in 
the form
\begin{equation}
  \Psi(t) \sim \sum_n \sum_{\Phi^{(n)}} \sum_{\Phi^{({\rm BH})}} 
  a_{\Phi^{(n)},\Phi^{({\rm BH})}} \Phi^{(n)} \Phi^{({\rm BH})},
\label{eq:ex-BH}
\end{equation}
where $\Phi^{({\rm BH})}$ are operators that cannot be obtained by OPEs 
of single-particle operators within the regime in which the particle 
interpretation is available, and $\Phi^{(n)}$ are (generalized) $n$-particle 
operators defined on the ``background'' of $\Phi^{({\rm BH})}$.

Since $\Phi^{({\rm BH})}$'s do not obey the property characterizing 
multi-particle operators, which is responsible for the emergence of the 
gravitational bulk picture, there is no reason to expect that the dynamics 
of these degrees of freedom (i.e.\ how $\Phi^{({\rm BH})}$'s respond to $D$) 
have a simple interpretation in the weakly coupled gravitational theory 
(although small components of them, labeled by $a$ in Eq.~(\ref{eq:A_i}), 
admit a semiclassical interpretation).  The spacetime attribute of 
these degrees of freedom, $k$, is defined only through interactions 
with the multi-particle degrees of freedom, $a^{\rm far}$ and $a$, 
as in Eqs.~(\ref{eq:Hawking-micro-2},~\ref{eq:mining-micro}).  In the 
distant description, this exhibits a thermodynamic characteristic given 
by Eq.~(\ref{eq:T-thermal}).  As argued in Refs.~\cite{Nomura:2014woa,%
Nomura:2014voa}, however, this attribution is expected to depend on 
the description (reference frame) one adopts.  Since the internal 
dynamics of the vacuum degrees of freedom does not have a simple 
spacetime interpretation in the gravitational bulk, the transfer 
of black hole microscopic information cannot be viewed as occurring 
through field (or string) theoretic modes on a fixed semiclassical 
background.  This was a crucial ingredient in the picture of 
Refs.~\cite{Nomura:2014woa,Nomura:2014voa} to avoid the firewall 
argument in Refs.~\cite{Almheiri:2012rt,Almheiri:2013hfa}.

The vacuum degrees of freedom naturally live in ($d$-dimensional) 
holographic spacetime.  The fact that their dynamics does not have a 
simple interpretation in the gravitational bulk does not mean that it is 
totally random:\ it still obeys constraints imposed by a $d$-dimensional 
CFT, which is a local field theory (though not on a spacetime on which 
the gravitational picture is built).  In particular, we expect that 
these degrees of freedom are classicalized in the $(d-1)$-dimensional 
space as time passes.  This suggests that, in the (fictitious) 
limit in which interactions with other systems are turned off, 
these degrees of freedom are subject to the classical Poincar\'{e} 
recurrence time $e^S$, not the quantum Poincar\'{e} recurrence time 
$e^{e^S}$ (see Ref.~\cite{Nomura:2015zda} for a related discussion).

\section{Summary and Discussion}
\label{sec:discuss}

If a global spacetime description of quantum gravity exists, it will be 
highly redundant.  In addition to standard diffeomorphism invariance, 
we expect redundancies associated with holography~\cite{'tHooft:1993gx,%
Susskind:1994vu} and large-scale causal structures~\cite{Susskind:1993if,%
Susskind:2005js}.  However, there certainly must be quantities free from 
these redundancies, such as causal relations among events.  How do we 
extract such physical information?

AdS/CFT duality provides a way to fix these redundancies for spacetimes 
that are asymptotically AdS:\ a unitary theory with conformal symmetry 
describes quantum gravity in these spacetimes.  This implies that the 
physical content of such theories is fully encoded in the algebra of 
operators ${\cal O}_I$'s which are representations of the conformal 
group.  While these operators are often combined to form fields, this 
need not be the case---they can all be viewed as located at the point 
$x = x_{-\infty}$.  Moreover, if the theory satisfies certain additional 
criteria, such as a large central charge, then the dual gravitational 
description possesses a large approximately flat spacetime region. 
In this case, a suitable (tiny) subset of operators ${\cal O}_I$'s 
contains virtually all of the information about flat-space quantum 
gravity (except for that associated with the global or large-time 
structure of the theory).  In particular, this set of operators, 
$\Psi_A$, can represent the action of sufficiently small dilatations; 
namely, $e^{-i D t} \Psi_A e^{i D t}$ for generic $A$ can be written 
as a superposition of $\Psi_A$'s with extremely high accuracy for 
$t \lesssim R$.  This allows us to describe continuous time evolution 
of processes in flat-space quantum gravity.

To physically interpret such processes, we need a way to decompose 
operators into smaller elements, i.e.\ the Hilbert space of the states 
for the entire universe into those for smaller subsystems.  This can 
be done by the OPE, defined by the action of $\Psi_A$'s on general 
states.  For configurations that do not have strong gravitational 
effects, this allows us to interpret $\Psi_A$ as multi-particle operators
\begin{equation}
  \Psi_A \sim \sum_n \sum_{\Phi^{(n)}} a_{\Phi^{(n)}} \Phi^{(n)},
\qquad
  \Phi^{(n)} \sim \left\{ \begin{array}{ll} 
    {\bf 1} & \mbox{for } n = 0,
  \vspace{2mm} \\
    \Phi_1^{(1)} \cdots \Phi_n^{(1)} & \mbox{for } n = 1,2,\cdots,
  \end{array} \right.
\end{equation}
where $\Phi_i^{(1)}$ are single-particle operators which can be defined 
at the nonperturbative level by their characteristic OPE structure. 
However, for configurations with strong gravitational effects, such as 
those with a black hole, this decomposition is not possible.  We can 
only write them as
\begin{equation}
  \Psi_A \sim \sum_n \sum_{\Phi^{(n)}} \sum_{\Phi^{({\rm BH})}} 
  a_{\Phi^{(n)},\Phi^{({\rm BH})}} \Phi^{(n)} \Phi^{({\rm BH})},
\end{equation}
where $\Phi^{({\rm BH})}$ cannot be obtained by OPEs of $\Phi_i^{(1)}$'s 
within the regime in which the particle interpretation is available.  We 
have argued that in the near black hole region, the degrees of freedom 
corresponding to $\Phi^{({\rm BH})}$ (labeled by $k$) completely dominate 
over those corresponding to $\Phi^{(n)}$ (labeled by $a$).  These 
dominant degrees of freedom $k$---called the vacuum degrees of freedom 
and responsible for most of the Bekenstein-Hawking entropy---do not obey 
simple semiclassical dynamics in the gravitational bulk.  Their spacetime 
attribute is defined only through their interactions with multi-particle 
operators. This is the essence of what was called spacetime-matter 
duality in Refs.~\cite{Nomura:2014woa,Nomura:2014voa}, a crucial 
element to make the existence of the black hole interior consistent 
with unitary evolution.

It is useful to describe explicitly how the construction here avoids 
the firewall argument in Ref.~\cite{Marolf:2013dba}, specifically 
tailored to the context of gauge/gravity duality.  Our assertion is 
that the black hole microstates represented by the Bekenstein-Hawking 
entropy correspond to states generated by operators $\Phi^{({\rm BH})}$. 
On the other hand, operators representing semiclassical excitations 
are associated with $\Phi^{(n)}$.  When (exterior) quantum field theory 
operators, $b$ and $b^\dagger$, are defined together with the Fock 
space generated by them, all of the black hole microstates are mapped 
into a particular thermal state (so that these microstates cannot be 
resolved at the field theory level, manifesting the coarse-graining 
performed to obtain the semiclassical picture).  Furthermore, physical 
excitations of the semiclassical degrees of freedom give only a slight 
perturbation to this overall picture.  This implies that the space of 
physical states is only a tiny portion of the naive Fock space spanned 
by all the possible states obtained by acting $b^\dagger$'s on the $b$ 
vacuum.  Therefore, the average over all the $b$ eigenstates employed 
in Ref.~\cite{Marolf:2013dba} is irrelevant to see properties of physical 
states, all of which are in (very) special linear combinations of the 
$b$ eigenstates.

We stress that while our scheme allows for describing the microscopic 
dynamics of an evaporating black hole using a CFT, and hence can test 
aspects of the proposal in Refs.~\cite{Nomura:2014woa,Nomura:2014voa}, 
the corresponding description in the bulk is the one viewed from the 
asymptotic infinity.  In particular, it does not guarantee that an 
infalling description in the bulk, which involves interior spacetime 
of the black hole and in which the time evolution operator takes a simple 
time-independent form, must be obtained within the CFT.  It is possible 
that in order to represent a bulk reference frame change in a holographic 
theory, we need to enlarge the operator set beyond that of the CFT, 
so that complementarity transformations---and in particular, the time 
evolution operators for infalling reference frames---can only be represented 
with this enlarged operator set.  Note that in order for this picture 
to make sense, the number of independent CFT operators, $N_{\cal O}$, must 
be much smaller than $({\rm dim}\,{\cal H})^2$, where ${\cal H}$ is the 
CFT Hilbert space.  This is indeed the case because of the operator-state 
correspondence, which dictates $N_{\cal O} = {\rm dim}\,{\cal H}$. 
A simple way to state this is that the CFT description of the bulk 
spacetime may be the one in which the reference point is ``anchored'' 
to the asymptotic infinity region.%
\footnote{In the Schr\"{o}dinger picture language adopted in 
 Refs.~\cite{Nomura:2011rb,Nomura:2013nya}, this is equivalent to saying 
 that Hilbert subspaces representing the system as viewed from an infalling 
 reference frame are not contained in the CFT Hilbert space.}
We stress that this is not an indication that the CFT can provide only 
an incomplete description of the bulk physics, nor does this prove the 
existence of firewalls.  It simply says that the CFT description arises 
only after the ``reference frame gauge'' has been fixed to an exterior---in 
fact, the asymptotically distant---one.%
\footnote{An alternative possibility is that the CFT formulation has 
 a freedom in choosing a reference frame in the bulk.  If an infalling 
 description is available in this way, it would have its own time evolution 
 operator and the basis states $\check{\cal B} = \{ \check{\Psi}_A \}$ 
 within the CFT, with the map ${\cal B} \leftrightarrow \check{\cal B}$ 
 defining the complementarity transformation.  While we do not know 
 if such a realization is indeed possible, if so, we would expect it 
 selects its own semiclassical operators $\{ \check{\Phi}^{(n)} \} 
 \subset \check{\cal B}$ and, as in the case of the distant description, 
 the majority of the degrees of freedom in the near black hole region 
 is represented by operators $\check{\Phi}^{({\rm BH})}$ that cannot be 
 interpreted as semiclassical multi-particle operators.  The spacetime 
 attribute of these operators would then be given only through their 
 interactions with $\check{\Phi}^{(n)}$'s, which would exhibit the property 
 called extreme relativeness in Refs.~\cite{Nomura:2014woa,Nomura:2014voa}.}

In summary, although current theoretical technology does not 
yet allow us to explicitly compute the microscopic dynamics 
of black holes, it is reassuring that a consistent picture 
exists~\cite{Nomura:2014woa,Nomura:2014voa} and that properties 
of the class of CFTs with weakly coupled gravitational descriptions 
can test (at least the distant description part of) it.  Hopefully 
further theoretical advancements will confirm these properties.

\section*{Acknowledgments}

We would like to thank Yu Nakayama, Yuji Tachikawa, Tadashi Takayanagi, 
and Taizan Watari for useful conversations on this and related topics. 
We also thank Kavli Institute for the Physics and Mathematics of the 
Universe, University of Tokyo for hospitality during the visit in which 
a part of this work was carried out.  This work was supported in part 
by the Director, Office of Science, Office of High Energy and Nuclear 
Physics, of the U.S.\ Department of Energy (DOE) under Contract 
DE-AC02-05CH11231, by the National Science Foundation under grants 
PHY-1214644 and PHY-1521446, and by MEXT KAKENHI Grant Number 15H05895. 
The work of F.S. was supported in part by the DOE National Nuclear 
Security Administration Stewardship Science Graduate Fellowship.

\end{document}